\documentclass[aps,twocolumn,floatfix]{revtex4-1}
\usepackage{bm} 
\usepackage{dcolumn} 
\newcommand{\RNum}[1]{\uppercase\expandafter{\romannumeral #1\relax}}
\usepackage{graphicx} 
\usepackage{amsmath} 
\usepackage{amssymb} 
\usepackage{amsfonts} 
\usepackage{physics}

\begin{document}
\title{Tunable quantum interference effect on magnetoconductivity \\
in few-layer black phosphorus}
\author{Sunghoon Kim}
\author{Hongki Min}
\email{hmin@snu.ac.kr}
\affiliation{Department of Physics and Astronomy, Seoul National University, Seoul 08826, Korea}
\date{\today}

\begin{abstract}
In this study, we develop a systematic weak localization and antilocalization theory fully considering the anisotropy and Berry phase of the system, and apply it to various phases of few-layer black phosphorus (BP), which has a highly anisotropic electronic structure with an electronic gap size tunable even to a negative value.
The derivation of a Cooperon ansatz for the Bethe-Salpeter equation in a general anisotropic system is presented, revealing the existence of various quantum interference effects in different phases of few-layer BP, including a crossover from weak localization to antilocalization. We also predict that the magnetoconductivity at the semi-Dirac transition point will exhibit a nontrivial power-law dependence on the magnetic field, while following the conventional logarithmic field dependence of two-dimensional systems in the insulator and Dirac semimetal phases. Notably, the ratio between the magnetoconductivity and Boltzmann conductivity turns out to be independent of the direction, even in strongly anisotropic systems. Finally, we discuss the tunability of the quantum corrections of few-layer BP in terms of the symmetry class of the system.
\end{abstract}

\maketitle

{\em Introduction.} --- The presence of disorder enriches the physics of mesoscopic systems \cite{Anderson1958, Lee1985, Datta1995, Akkermans2007}. One of the remarkable effects of disorder on transport properties is a negative correction to the dc conductivity, so called weak localization (WL). WL arises as electrons obtain an enhanced probability of backscattering due to a quantum interference effect. The theoretical description of the WL effect is accessible in a Feynman diagram formalism \cite{Hikami1980, Altshuler1980, Wolfle1984, Beutler1988}, and its signal has been explored extensively through experiments \cite{Bergmann1982, Bergmann1984}. In contrast, 
electrons carrying a \(\pi\) Berry phase result in a positive quantum correction to the dc conductivity, 
which is referred to as weak antilocalization (WAL). 
WAL is attributed to a suppressed probability of backscattering due to an additional minus sign by the \(\pi\) Berry phase.
The WAL effect has been studied intensely in topological states of matter with the \(\pi\) Berry phase, such as graphene \cite{Suzuura2002, McCann2006, Gorbachev2007, Kechedzhi2007, Nomura2007, Imura2009}, topological semimetals \cite{Lu2015, Dai2016, Lu2017, Chen2019, Fu2019}, and surface states of topological insulators \cite{He2011, Lu2011, Zhang2012}. Such materials show complex quantum interference corrections due to the competition between the WL and WAL effects \cite{Suzuura2002, McCann2006, Fu2019}.

\begin{figure}[htb]
\includegraphics[width=1\linewidth,height=2.3in]{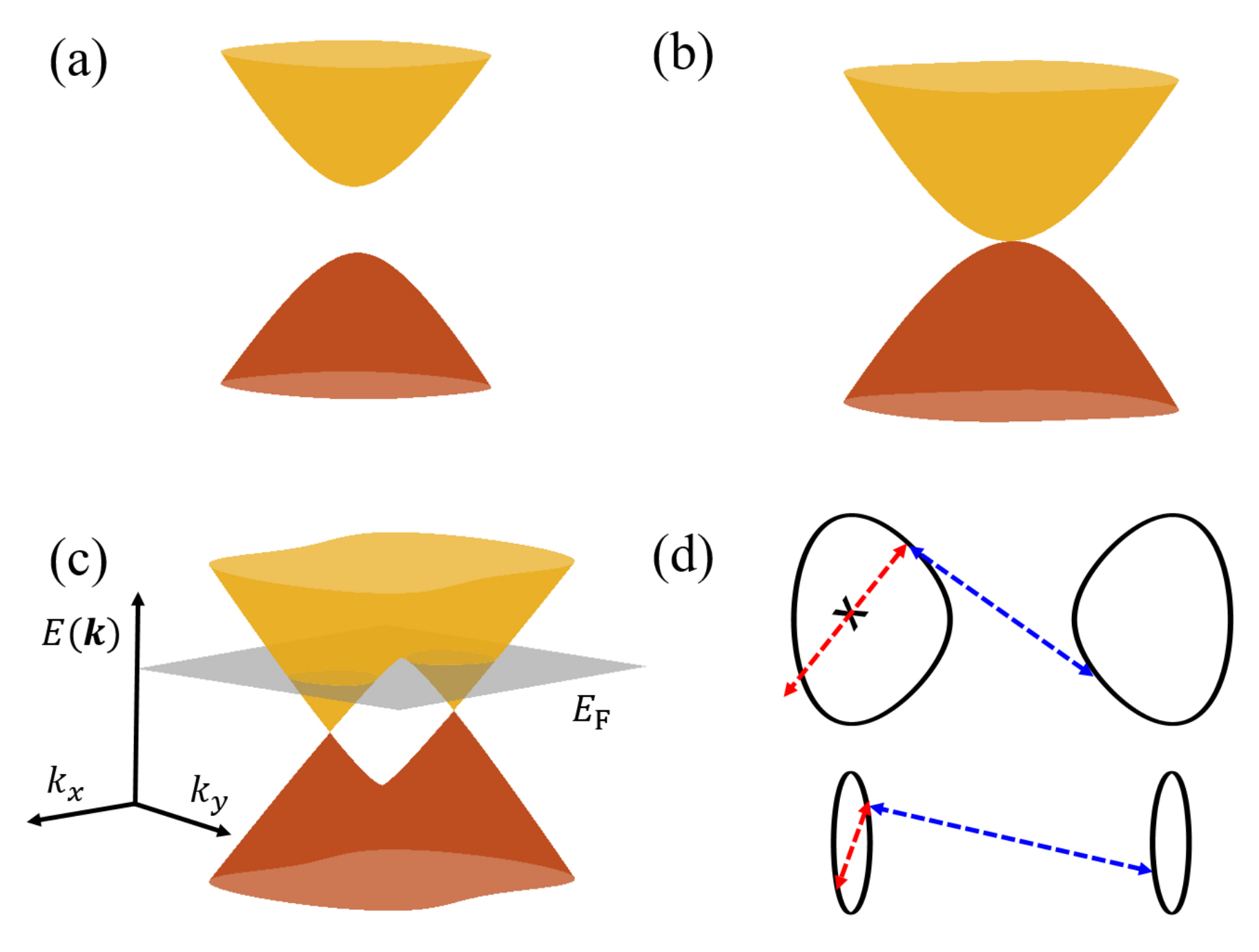}
\caption{
Electronic structure of few-layer BP in the (a) insulator phase, (b) semi-Dirac transition point (SDTP), and (c) Dirac semimetal (DSM) phase. (d) The Fermi surfaces of the DSM phase. At a sufficiently low Fermi energy (lower panel), a momentum state has time-reversed counterparts
both within the node and in the opposite node. Thus, the quantum interference effect is contributed by two types of scatterings: intranode (red dashed arrow) and internode (blue dashed arrow) scatterings. As the Fermi energy increases (upper panel), the Fermi surface is distorted and the time-reversal symmetry around a node is broken. Thus, the quantum interference effect via intranode scattering is suppressed. }
\label{fig:1}
\end{figure}

Few-layer black phosphorus (BP), a two-dimensional (2D) van der Waals material which has been studied intensely both theoretically \cite{Rodin2014, Rudenko2014, Liu2015, Yuan2015, Kim2015, Pereira2015, Baik2015, Adroguer2016, Yuan2016, Doh2017, Kim2017, Park2019, Jang2019} and experimentally \cite{Li2014, Xia2014, Tran2014, Qiao2014, Xiang2015, Du2016, Shi2016, Hemsworth2016, Li2017}, is expected to show nontrivial quantum interference effects due to strong anisotropy and a tunable electronic structure. Few-layer BP has a direct band gap [Fig.~\ref{fig:1}(a)], which can be tuned by external perturbations, such as strain \cite{Rodin2014}, pressure \cite{Xiang2015}, electric gating \cite{Liu2015, Yuan2016}, and chemical doping \cite{Baik2015, Kim2015}. Such modulations can close the band gap, resulting in a semi-Dirac transition point (SDTP) 
with a combination of linear and quadratic dispersions [Fig.~\ref{fig:1}(b)]. These modulations can even induce a band inversion, leading to the Dirac semimetal (DSM) phase with linear dispersions around the nodes [Fig.~\ref{fig:1}(c)].
Although there have been a few experimental studies on the quantum interference effects in few-layer BP \cite{Du2016, Shi2016, Hemsworth2016, Li2017}, a theoretical approach on each phase has been elusive due to its nontrivial anisotropy, which cannot be described by a simple model with different effective masses in each direction. Thus, a further systematic formalism of the quantum interference theory is called for. 

In this study, we develop a general framework for the quantum interference effect, which captures the anisotropy and Berry phase effect of the system. This is achieved by deriving a Cooperon ansatz which applies to a general system with an anisotropic energy dispersion. By applying this framework to the various phases of few-layer BP, 
we find that the insulator phase and SDTP have the anisotropic WL corrections, while the DSM phase can host either WL or WAL arising from internode and intranode scatterings, respectively. 
We also estimate the effect of a magnetic field on the quantum corrections in each phase, predicting a nontrivial power-law dependence on the magnetic field at the SDTP. Notably, we show that the ratio between the magnetoconductivity and Boltzmann conductivity is independent of the direction, even in highly anisotropic systems. Finally, we discuss the tunability of the quantum interference effect in few-layer BP, with comments on the symmetry class.

{\em Model.} --- The low-energy effective Hamiltonian of few-layer BP is given by \cite{Baik2015, Doh2017, Park2019, Jang2019}
\begin{equation}
\label{eq:H}
    H=\left(\frac{\hbar^{2}k^{2}_{x}}{2m^{*}}+\frac{E_{\rm{g}}}{2}\right)\sigma_{x}+\hbar v_{y}k_{y}\sigma_{y},
\end{equation}
where \(m^{*}\) is the effective mass along the zigzag ($x$) direction, \(E_{\rm{g}}\) is the band gap, \(v_{y}\) is the velocity along the armchair ($y$) direction, and \(\sigma_{x,y}\) are the Pauli matrices. The corresponding energy eigenvalues are given by \(E=\pm \sqrt{\left(\frac{\hbar^{2}k^{2}_{x}}{2m^{*}}+\frac{E_{\rm{g}}}{2}\right)^{2}+\hbar^{2}v^{2}_{y}k_{y}^{2}}\). Without band gap tuning, the Hamiltonian has a direct band gap \((E_{\rm{g}}>0)\) and few-layer BP is in the gapped insulator phase [Fig.~\ref{fig:1}(a)]. At \(E_{\rm{g}}=0\), the Hamiltonian has a band touching point at \((k_{x},k_{y})=(0,0)\) and few-layer BP is in the SDTP [Fig.~\ref{fig:1}(b)]. At this point, the energy dispersion is linear in the armchair direction and quadratic in the zigzag direction. As the band gap decreases further to a negative value \((E_{\rm{g}}<0)\), few-layer BP becomes the DSM phase with two nodes at \(\bm{K}^{\pm}=(\pm \sqrt{\frac{m^{*}|E_{\rm{g}}|}{\hbar^{2}
}},0)\) [Fig.~\ref{fig:1}(c)]. At a sufficiently low Fermi energy satisfying \(E_{\rm{F}}\ll |E_{\rm{g}}|\), the Hamiltonian near each node can be linearized as
\begin{eqnarray}
\label{eq:Hamiltonian_DSM}
H&=&\pm \hbar v_{x}\left(k_{x}-K^{\pm}_{x}\right) \sigma_{x}+ \hbar v_{y}k_{y}\sigma_{y} \nonumber \\
&\equiv&\hbar v_{0}\kappa(\pm \cos{\phi_{\bm{\kappa}}}\sigma_{x}+\sin{\phi_{\bm{\kappa}}}\sigma_{y}),
\end{eqnarray}
where \(v_{x}=\sqrt{\frac{|E_{\rm{g}}|}{m^{*}}}\) is the velocity along the zigzag direction. For later convenience,
we adopt the parametrization \(v_{0}\kappa\cos{\phi_{\bm{\kappa}}}=v_{x}\left(k_{x}-K^{\pm}_{x}\right)\) and \(v_{0}\kappa\sin{\phi_{\bm{\kappa}}}=v_{y}k_{y}\), where \(\bm{\kappa}\) is an effective momentum measured from the nodes. 
In this work, we neglect the effects of spin-orbit coupling due to its negligible size in few-layer BP.


The quantum correction is determined not only by the electronic structure, but also by the type of impurity. The quantum interference effect is only induced by static and non-magnetic impurities as it is destroyed by non-static or magnetic impurities \cite{Datta1995, Akkermans2007}. Another aspect we should consider is the range of scattering. At low Fermi energies, regardless of the range of scattering, an electron in the insulator phase and SDTP has a single time-reversed counterpart on the whole Fermi surface. In contrast, for the DSM phase we consider two types of scatterings: intranode and internode scatterings [Fig. ~\ref{fig:1}(d)]. In the \(E_{\rm{F}}\ll E_{\rm{g}}\) limit, intranode scattering may occur by long-range impurities, whereas internode scattering may arise from short-range impurities, such as lattice vacancies. The relative strengths of intranode and internode scatterings lead to competition between the WAL and WL effects in the DSM phase, as will be discussed later.

\begin{figure}[htb]
\includegraphics[width=1\linewidth,height=2.4in]{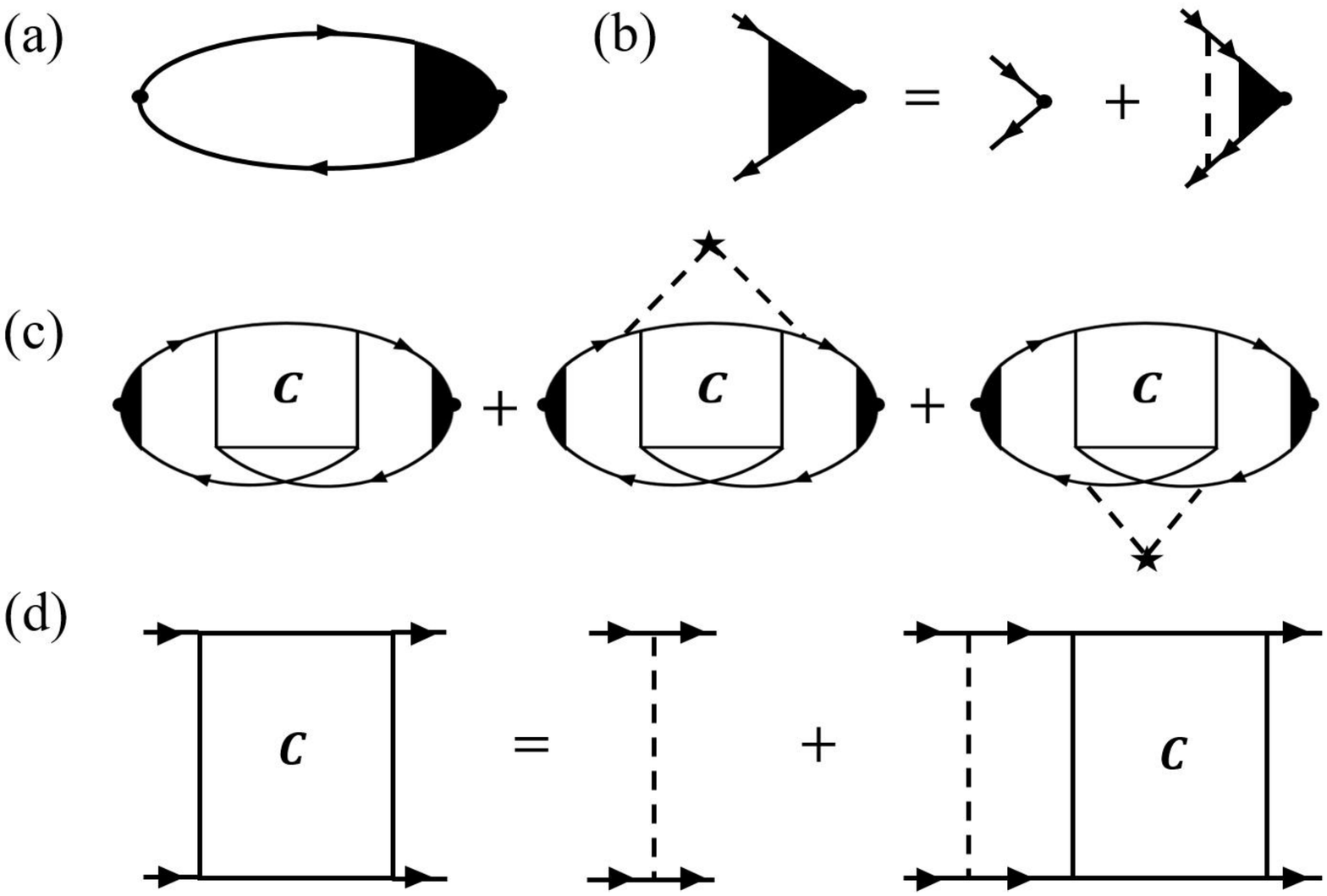}
\caption{
Feynman diagrams describing the corrections to the dc conductivity. (a) The current-current correlation function supplemented with the ladder vertex correction gives results equivalent to the Boltzmann transport theory. (b) The ladder vertex correction satisfies the self-consistent Dyson's equation. (c) The quantum correction to the dc conductivity is mostly contributed by a bare Hikami box and two dressed Hikami boxes. (d) The Cooperon operator obeys the self-consistent Bethe-Salpeter equation. 
}
\label{fig:2}
\end{figure}


{\em Diagrammatic approach.} --- In a diagrammatic approach, effects of weak disorder on transport can be studied by incorporating relevant corrections into the current-current correlation function with disorder-averaged Green's functions \cite{Mahan2000, Coleman2016}. Calculating the self-energy up to the first-order Born approximation, we obtain the quasiparticle lifetime $(\tau^{\rm{qp}}_{\bm{k}})^{-1}={1\over\mathcal{V}}\sum_{\bm{k}'}W_{\bm{k}',\bm{k}}$ where $\mathcal{V}$ is the volume of the system and $W_{\bm{k}',\bm{k}}$ is the transition rate from $\bm{k}$ to $\bm{k}'$. Note that for impurity scattering, the transition rate is given by $W_{\bm{k}',\bm{k}}={2\pi n_{\rm{imp}}\over\hbar}|V_{\bm{k}',\bm{k}}|^{2}\delta(\xi_{\bm{k}}-\xi_{\bm{k}'})$ where $n_{\rm{imp}}$ is the impurity density, $V_{\bm{k'},\bm{k}}=\matrixel{\bm{k}'}{V}{\bm{k}}$ is the matrix element of the scattering potential $V$, and $\xi_{\bm{k}}\equiv \varepsilon_{\bm{k}}-\mu$ is the energy of an electron with respect to the chemical potential \(\mu\). 
The ladder vertex correction combined with the first-order Born approximation is the leading impurity correction to the current vertex [Fig.~\ref{fig:2}(a)], which gives results equivalent to those of the semiclassical Boltzmann transport theory. The ladder vertex correction obeys the self-consistent Dyson's equation [Fig.~\ref{fig:2}(b)], and for isotropic systems, it gives a well-known result with the $(1-\cos\theta)$ factor in the inverse transport relaxation time \cite{Ashcroft1976}, suppressing the low-angle scattering contribution to transport. In an anisotropic system, the relation is generalized to a coupled integral equation, which relates transport relaxation times at different states. In a \(d\)-dimensional anisotropic system, it reads \cite{Park2017, Park2019, Kim2019}

\begin{equation}
\label{eq:coupled integral equation}
    1=\int \frac{d^{d}k'}{(2\pi)^d}W_{\bm{k}',\bm{k}}\left(\tau^{(i)}_{\bm{k}}-\frac{v^{(i)}_{\bm{k'}}}{v^{(i)}_{\bm{k}}}\tau^{(i)}_{\bm{k'}}\right),
\end{equation}
where \(\tau^{(i)}_{\bm{k}}\) and \(v^{(i)}_{\bm{k}}=\frac{1}{\hbar}\frac{\partial \varepsilon _{\bm{k}}}{\partial k_{i}}\) are the transport relaxation time and velocity, respectively, along the \(i\)th direction.
The ladder vertex correction modifies the velocity vertex along the \(i\)th direction as \(\tilde{v}^{(i)}_{\bm{k}}=v^{(i)}_{\bm{k}}\tau^{(i)}_{\bm{k}}
/\tau^{\rm{qp}}_{\bm{k}}\) \cite{Kim2019}. In the zero-temperature limit, the ladder approximation yields the dc conductivity in the semiclassical Boltzmann regime as
\begin{eqnarray}
\label{eq:Boltzmann conductivity}
\sigma_{ij}^{\rm{B}}=ge^{2} \int \frac{d^{d}k}{(2\pi)^d} \delta(\xi_{\bm{k}}) v_{\bm{k}}^{(i)} v_{\bm{k}}^{(j)} \tau_{\bm{k}}^{(j)},
\end{eqnarray}
where \(g\) is the degeneracy of the system and the superscript \(\rm{B}\) denotes the Boltzmann conductivity.

Consideration of further diagrams, called maximally crossed diagrams, leads to the quantum interference correction [Figs.~\ref{fig:2}(c) and ~\ref{fig:2}(d)]. The quantum correction can be boiled down to three leading terms referred to as a bare Hikami box and two dressed Hikami boxes \cite{McCann2006}. The Hikami boxes can be computed by the Cooperon operator, which obeys the following Bethe-Salpeter equation \cite{Datta1995,Akkermans2007} (in the following, we omit \(\hbar\) for simplicity):
\begin{eqnarray}
\label{eq:bethe-salpeter_main1}
C^{\rm{AR}}_{\bm{Q}}(\bm{k},\bm{k}')&&=n_{\rm{imp}}V_{\bm{k'},\bm{k}}V_{-\bm{k'},-\bm{k}}
\nonumber\\&&+\frac{n_{\rm{imp}}}{\mathcal{V}}\sum_{\bm{p}}V_{\bm{p},\bm{k}}V_{\bm{-p},\bm{-k}}C^{\rm{AR}}_{\bm{Q}}(\bm{p},\bm{k}')
\nonumber\\&&\times G^{\rm{A}}(\bm{p},0)G^{\rm{R}}(\bm{Q}-\bm{p},0),
\end{eqnarray}
where \(C^{\rm{AR}}_{\bm{Q}}(\bm{k},\bm{k}')\) is the Cooperon with the momentum \(\bm{Q}=\bm{k}+\bm{k}'\), and the superscripts \(\rm{A}\) and \(\rm{R}\) represent the advanced and retarded functions, respectively. Note that frequencies in the Cooperon are set to zero as we focus on the dc conductivity. 
The retarded and advanced Green's functions are \(G^{\rm{R,A}}(\bm{p},\xi)=[\xi-\xi_{\bm{p}}\pm i/2\tau_{\bm{p}}^{\rm{qp}}]^{-1}\). As the Green's functions near the Fermi surface contribute mostly to the momentum summation in the right-hand side of Eq.~(\ref{eq:bethe-salpeter_main1}), we can perform the \(\xi_{\bm{p}}-\)integral separately as  \(\int d\xi_{\bm{p}} G^{\rm{A}}(\bm{p},0)G^{\rm{R}}(\bm{Q}-\bm{p},0)=2\pi i\left[\bm{Q}\cdot \bm{v}_{\bm{p}}+\frac{i}{2}\left(\frac{1}{\tau^{\rm{qp}}_{\bm{p}}}+\frac{1}{\tau^{\rm{qp}}_{\bm{p}-\bm{Q}}}\right)\right]^{-1}\) \cite{Coleman2016}. Using the Ward identities in anisotropic systems \cite{Kim2019}, we rewrite the denominator of the right-hand side of the \(\xi_{\bm{p}}-\)integral as
\begin{eqnarray}
\bm{Q}\cdot \bm{v}_{\bm{p}}+\frac{i}{2}\left(\frac{1}{\tau^{\rm{qp}}_{\bm{p}}}+\frac{1}{\tau^{\rm{qp}}_{\bm{p}-\bm{Q}}}\right)=\frac{1}{\tau^{\rm{qp}}_{\bm{p}}}\left[i+f_{\bm{Q}}(\bm{p})\right],
\end{eqnarray}
where \(f_{\bm{Q}}(\bm{p})\equiv \sum_{i}Q^{(i)}v^{(i)}_{\bm{p}}\tau^{(i)}_{\bm{p}}\). Accordingly, we have the Bethe-Salpeter equation in anisotropic systems as follows: 
\begin{eqnarray}
\label{eq:bethe-salpeter_main2}
C^{\rm{AR}}_{\bm{Q}}(\bm{k},\bm{k}')
&&\approx n_{\rm{imp}}V_{\bm{k'},\bm{k}}V_{-\bm{k'},-\bm{k}}\nonumber\\&&+\frac{2\pi n_{\rm{imp}}}{\mathcal{V}}\sum_{\bm{p}}\delta(\xi_{\bm{p}})V_{\bm{p},\bm{k}}V_{-\bm{p},-\bm{k}}C^{\rm{AR}}_{\bm{Q}}(\bm{p},\bm{k}')
\nonumber\\&&\times \tau^{\rm{qp}}_{\bm{p}}\left[1+if_{\bm{Q}}(\bm{p})-f_{\bm{Q}}^{2}(\bm{p})\right],
\end{eqnarray}
where the Cooperon diverges as \(Q\rightarrow 0\), and thus we ignore terms of order higher than \(Q^{2}\). For the detailed derivations, see the Supplemental Material \cite{see SM}. Importantly, we capture the full anisotropy of the system by introducing \(f_{\bm{Q}}(\bm{p})\), which is determined by the anisotropic velocities and transport relaxation times on the Fermi surface. 

The Bethe-Salpeter equation in Eq.~(\ref{eq:bethe-salpeter_main2}) can be solved self-consistently by collecting the most divergent contribution in the \(Q\rightarrow 0\) limit \cite{Suzuura2002,Lu2015,Chen2019}. Alternatively, we directly derive the following Cooperon ansatz by computing the average of Eq.~(\ref{eq:bethe-salpeter_main2}) on the Fermi surface \cite{see SM}:
\begin{eqnarray}
\label{eq:ansatz}
C^{\rm{AR}}_{\bm{Q}}(\bm{k},\bm{k'})=\frac{(2\pi N_{0}\tau^{\rm{qp}}_{\bm{k}}\tau^{\rm{qp}}_{\bm{k'}})^{-1}}{\sum_{i,j}D_{ij}Q_{i}Q_{j}}F(\bm{k},\bm{k}'),
\end{eqnarray}
where \(N_{0}\) is the density of states at the Fermi surface and \(D_{ij}\) are the diffusion coefficients defined by
\begin{eqnarray}
\label{diffusion tensor}
D_{ij}=\frac{1}{N_{0}\mathcal{V}}\sum_{\bm{p}}\delta(\xi_{\bm{p}})v_{\bm{p}}^{(i)}v_{\bm{p}}^{(j)}\tau^{(i)}_{\bm{p}}\tau^{(j)}_{\bm{p}}(\tau^{\rm{qp}}_{\bm{p}})^{-1}.
\end{eqnarray}
It is worth noting that the ansatz in Eq.~(\ref{eq:ansatz}) includes the electronic structure-dependent phase factor \(F(\bm{k},\bm{k}')\) which is defined via \(V_{\bm{-k}',\bm{-k}}\equiv V^{*}_{\bm{k'},\bm{k}} F(\bm{k},\bm{k}')\) \cite{see SM}. 
The ansatz is valid as long as the composition property \(F(\bm{k}_{1},\bm{k}_{3})=F(\bm{k}_{1},\bm{k}_{2})F(\bm{k}_{2},\bm{k}_{3})\) holds, as in the case of the various phases of few-layer BP. For the insulator phase and SDTP, \(F(\bm{k},\bm{k}')\) is unity, whereas for the DSM phase, \(F(\bm{\kappa},\bm{\kappa}')=e^{i(\phi_{\bm{\kappa}}-\phi_{\bm{\kappa}'})}\) (unity) for intranode (internode) scattering. Notably, the phase factor reflects the Berry phase of the system which determines the sign of the quantum correction, as will be shown below.

{\em WL and WAL corrections.} --- Using the ansatz in Eq.~(\ref{eq:ansatz}), we compute the total quantum correction \(\Delta \sigma_{ii}=\Delta \sigma^{\rm{bare}}_{ii}+2\Delta \sigma^{\rm{dressed}}_{ii}\) for each phase \cite{see SM}. For the insulator phase and SDTP, we find the following WL correction: 
\begin{eqnarray}
\label{eq:insulator_hikami_total_main}
\frac{\Delta \sigma_{ii}}{\sigma_{ii}^{\rm{B}}}=-\frac{1}{2\pi^2 N_{0}D\hbar}\ln{\left(\frac{\ell_{\phi}}{\ell_{e}}\right)},
\end{eqnarray}
where \(\ell_{\phi}\) is the phase coherence length, \(\ell_{e}\) is the mean free path, and \(D\equiv \sqrt{D_{xx}D_{yy}}\). We note that Eq.~(\ref{eq:insulator_hikami_total_main}) applies not only to few-layer BP but also to a general 2D anisotropic system regardless of scattering potential \cite{see SM}, with the additional minus sign for the WAL correction. Remarkably, the ratio of the quantum correction to the Boltzmann conductivity is the same irrespective of direction even in strongly anisotropic systems, which also holds in 3D anisotropic systems \cite{see SM}.  

For the DSM phase, we compute the quantum corrections due to intranode and internode scatterings, respectively. 
In this study, for simplicity we consider constant intranode and internode scattering potentials given by \(V_{\rm{intra}}\) and \(V_{\rm{inter}}\), respectively. When backscattering is governed by intranode scattering, the Cooperon is given by 
\begin{eqnarray}
C^{\rm{AR}}_{\bm{Q}}(\bm{\kappa},\bm{\kappa}')=\frac{1}{4(\tau^{\rm{qp}}_{\rm{intra}})^{2}}\frac{n_{\rm{imp}}V^{2}_{\rm{intra}}e^{i(\phi_{\bm{\kappa}}-\phi_{\bm{\kappa}'})}}{v_{x}^{2}Q^{2}_{x}+v_{y}^{2}Q^{2}_{y}},
\end{eqnarray}
where \((\tau^{\rm{qp}}_{\rm{intra}})^{-1}=\frac{n_{\rm{imp}}V^{2}_{\rm{intra}}E_{\rm{F}}}{2v_{x}v_{y}}\). Importantly, the phase factor \(F(\bm{\kappa},\bm{\kappa}')=e^{i(\phi_{\bm{\kappa}}-\phi_{\bm{\kappa}'})}\) reflects the \(\pi\) Berry phase of the DSM phase, which yields the WAL correction to the dc conductivity as
\begin{eqnarray}
\label{eq:dirac_WAL_xx}
\frac{\Delta \sigma_{ii}}{\sigma_{ii}^{\rm{B}}}=\frac{\hbar}{4\pi E_{\rm{F}}\tau_{\rm{intra}}^{\rm{qp}}}\ln\left(\frac{\ell_{\phi}}{\ell_{\rm{intra}}}\right), 
\end{eqnarray}
where \(\ell_{\rm{intra}}\) is the mean free path for intranode scattering. On the other hand, the Cooperon for internode scattering is obtained as
\begin{eqnarray}
C^{\rm{AR}}_{\bm{Q}}(\bm{\kappa},\bm{\kappa}')=\frac{1}{4(\tau_{\rm{inter}}^{\rm{qp}})^{2}}\frac{n_{\rm{imp}}V^{2}_{\rm{inter}}}{v_{x}^{2}Q^{2}_{x}+v_{y}^{2}Q^{2}_{y}},
\end{eqnarray}
where \((\tau^{\rm{qp}}_{\rm{inter}})^{-1}=\frac{n_{\rm{imp}}V^{2}_{\rm{inter}}E_{\rm{F}}}{2v_{x}v_{y}}\).
Thus, internode scattering gives rise to the WL correction as 
\begin{eqnarray}
\label{eq:dirac_WL_xx}
\frac{\Delta \sigma_{ii}}{\sigma_{ii}^{\rm{B}}}=-\frac{\hbar}{4\pi E_{\rm{F}}\tau_{\rm{inter}}^{\rm{qp}}}\ln\left(\frac{\ell_{\phi}}{\ell_{\rm{inter}}}\right),
\end{eqnarray}
where \(\ell_{\rm{inter}}\) denotes the mean-free path for internode scattering.

\begin{figure}[htb]
\includegraphics[width=1\linewidth]{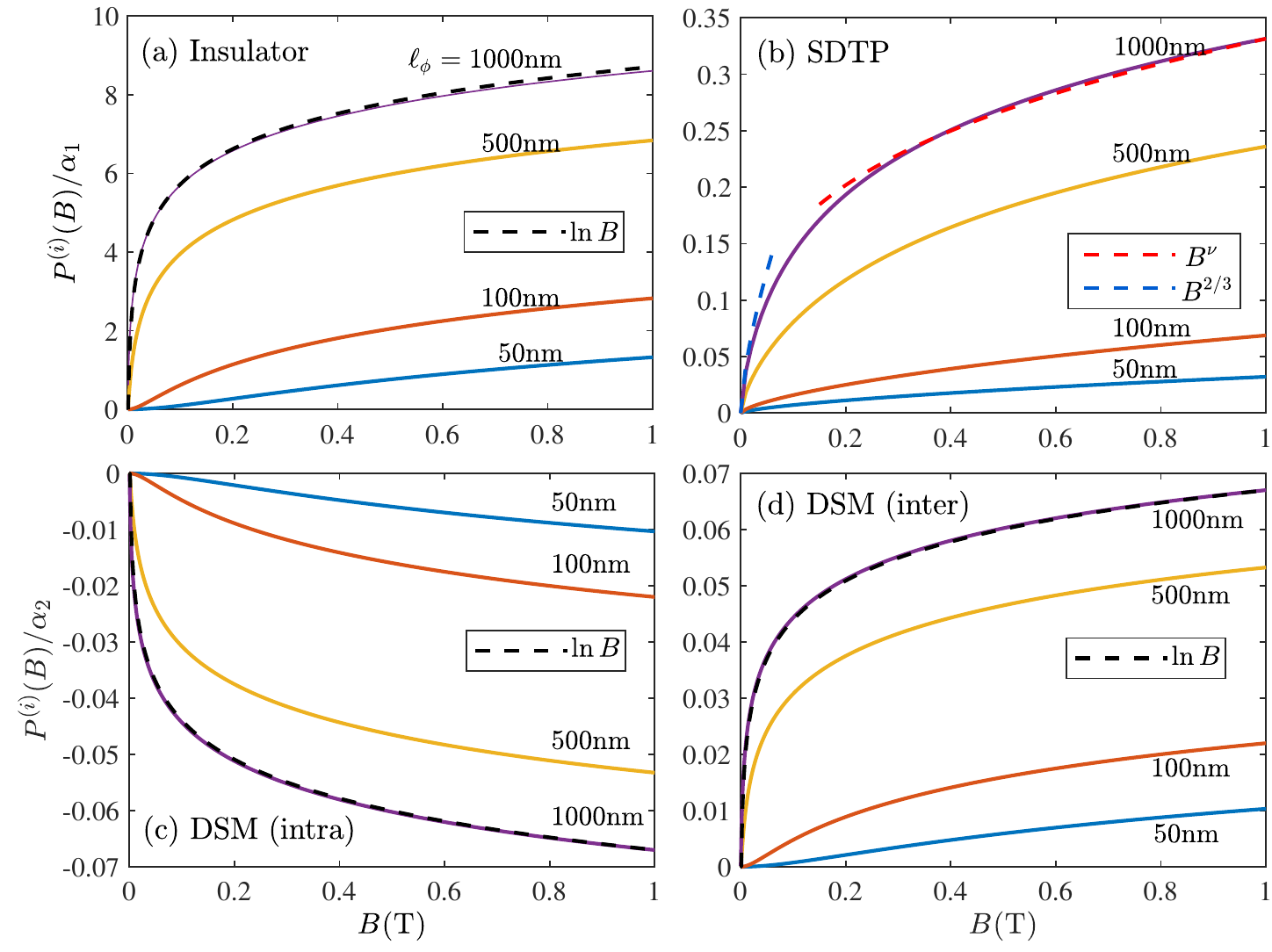}
\caption{
Ratio of the field-induced change in the magnetoconductivity to the Boltzmann conductivity for various phase coherence lengths with the mean free path \(\ell_{e}=10{\rm nm}\). The ratio for the
WL correction in (a) the insulator phase and (b) SDTP are plotted in units of \(\alpha_{1}=n_{\rm{imp}}V_{\rm{imp}}^{2}/\hbar^{2}v_{y}^{2}\), where \(V_{\rm{imp}}\) denotes the strength of the impurity potential. As for the DSM phase, (c) intranode scattering induces the WAL effect, (d) while internode scattering leads to the WL effect. Both corrections are plotted in units of \(\alpha_{2}=n_{\rm{imp}}V_{\rm{imp}}^{2}/\hbar^{2}v_{x}v_{y}\). As for the field dependence, the black dashed lines denote the \(\pm\ln{B}\) dependence, while the red dashed line and blue dashed line in (b) represent the \(B^{\nu}\) dependence with \(\nu \approx 0.31\) and \(B^{2/3}\) dependence, respectively. The exponent $\nu$ is not universal but depends on the system parameter.
}
\label{fig:3}
\end{figure}


{\em Magnetoconductivity.} --- Applying an external magnetic field gives an additional phase to each backscattering path, and thus destroys the quantum interference effect \cite{Lee1985, Datta1995, Akkermans2007}. In experiments, the phase coherence length can be obtained through the measurement of magnetoconductivity. The magnetoconductivity can be computed using the quantization of Landau levels \cite{Datta1995, Chen2019}. Assuming an external magnetic field along the \(z\) direction, we obtain the ratio 
\(P^{(i)}(B)\equiv[\Delta\sigma_{ii}(B)-\Delta\sigma_{ii}(0)]/\sigma^{\rm{B}}_{ii}
\) for each phase as follows \cite{see SM}:

\begin{subequations}
\begin{eqnarray}
\label{eq:magneto_insulator}
P^{(i)}_{\rm{ins}}(B)&=&-\frac{1}{4\pi^{2}N_{0}D\hbar}\left[\Psi\left(\frac{1}{2}+\frac{\ell^{2}_{B}}{\ell^{2}_{e}}\right)\right. \nonumber \\
&&\left.-\Psi\left(\frac{1}{2}+\frac{\ell^{2}_{B}}{\ell^{2}_{\phi}}\right)-2\ln{\left(\frac{\ell_{\phi}}{\ell_{e}}\right)}\right], \\
\label{eq:magneto_SDTP}
P^{(i)}_{\rm{SDTP}}(B)&=&\frac{1}{\pi^{3} D N_{0}\hbar}  \int_{\ell^{-1}_{\phi}}^{\ell^{-1}_{e}}d\tilde{Q}_{x}\frac{1}{\tilde{Q}_{x}} \nonumber \\
&&\times \tan^{-1}\left({\frac{\alpha^{-\frac{2}{3}}}{2^{\frac{2}{3}}\tilde{Q}_{x} \hbar v_{y}}}\sqrt{\frac{D_{yy}}{D}}\right),\\
\label{eq:magneto_DSM}
P^{(i)}_{\rm{DSM}}(B)&=&\frac{\hbar}{8\pi E_{\rm{F}}\tau_{\rm{intra}}^{\rm{qp}}} \left[\Psi\left(\frac{1}{2}+\frac{\ell^{2}_{B}}{\ell^{2}_{\rm{intra}}}\right)\right. \nonumber \\
&&\left.-\Psi\left(\frac{1}{2}+\frac{\ell^{2}_{B}}{\ell^{2}_{\phi}}\right)-2\ln{\left(\frac{\ell_{\phi}}{\ell_{\rm{intra}}}\right)}\right], 
\end{eqnarray}
\end{subequations}
where \(\Psi(x)\) is the digamma function, and \(\alpha\equiv\frac{\sqrt{m^{*}}}{v_{y}}\frac{2\ell_{B}^{2}}{\hbar^{2}A^{\frac{3}{2}}}\) with \(A \approx 1.17325\) \cite{Dietl2008} and \(\ell_{B}=\sqrt{\hbar c\over 4eB}\) is the magnetic length. Equation (\ref{eq:magneto_DSM}) is obtained for intranode scattering, and the field dependence for internode scattering only differs by the sign, with \(\ell_{\rm intra}\) (\(\tau^{\rm qp}_{\rm intra}\)) replaced by \(\ell_{\rm inter}\) (\(\tau^{\rm qp}_{\rm inter}\)). For the insulator phase [Fig.~\ref{fig:3}(a)] and DSM phase [Figs.~\ref{fig:3}(c) and  ~\ref{fig:3}(d)], we find that the field dependence of the magnetoconductivity is well approximated by \(P^{(i)}(B)\varpropto \pm \ln{B}\), which is in good agreement with the conventional prediction for 2D systems \cite{Hikami1980}. However, we predict that the magnetoconductivity of the SDTP will not follow the conventional prediction, but rather will show 
\(P^{(i)}(B)\varpropto B^{2\over 3}\) dependence in the weak field limit ( \(\ell_{\phi} \ll \ell_{B} \)), whereas \(P^{(i)}(B)\varpropto B^{\nu}\) dependence in the intermediate field regime (\(\ell_{e}\leq \ell_{B}\ll \ell_{\phi}\)) with the exponent \(\nu\), which depends on the system parameter and can be found numerically \cite{see SM}. 
This nontrivial field dependence may be attributed to strong anisotropy in the band dispersion \cite{Akkermans2007}, leading to a quantum diffusion which deviates from the 2D behavior. We note that \(P^{(i)}(B)\), the ratio of the field-induced change in the magnetoconductivity to the Boltzmann conductivity, is the same irrespective of the direction \(i\) for each phase.

{\em Discussion.} --- Our analysis shows that the sign of the quantum correction is determined by the electronic structure-dependent phase factor \(F(\bm{k},\bm{k}')\), which reflects the symmetry class of the system \cite{Suzuura2002, Lu2017, Dyson1962}. The unity phase factor of the insulator phase and SDTP indicates that the system has spinless time-reversal symmetry, belonging to the orthogonal class. Backscattering induced by a time-reversal operator of this kind leads to WL. As for the DSM phase, in the absence of internode scattering, the phase factor \(F(\bm{\kappa},\bm{\kappa}')=e^{i(\phi_{\bm{\kappa}}-\phi_{\bm{\kappa}'})}\) indicates that the system has time-reversal symmetry around a node without spin-rotational symmetry, belonging to the symplectic class. A system possessing a time-reversal operator of this kind exhibits WAL. In contrast, the presence of internode scattering can induce a crossover from the symplectic to orthogonal class, leading to the corresponding crossover from WAL to WL \cite{Suzuura2002}. Thus, the overall quantum correction in the DSM phase is determined by the dominant scattering mechanism, which depends on the separation between the two nodes \cite{see SM}. In the \(E_{F}\ll |E_{\rm{g}}|\) limit, a large separation between the nodes will suppress the internode scattering rates, and thus the WAL effect might be dominant over the WL effect. In addition, increasing the Fermi energy leads to the distortion of the Fermi surface, suppressing WAL.

Our results can be expanded to multiband systems straightforwardly. By including the interference contribution from the time-reversed paths in each band and associated Berry phase effect, we can compute the overall quantum corrections \cite{see SM}.

In this study, we developed a quantum interference theory for anisotropic systems by solving the Bethe-Salpeter equation for the Cooperon operator, fully considering the anisotropy and Berry phase of the system. 
We elaborated the Cooperon ansatz and diffusion coefficients in a compact and physically intuitive form with transport relaxation times in anisotropic systems, 
generalizing the previous work by P. W\"olfle {\em et al.} \cite{Wolfle1984}.
Furthermore, we considered systems with a nontrivial Berry phase, providing a systematic quantum interference theory for both WL and WAL effects, and the crossover between them. 

\acknowledgments
This work was supported by the National Research Foundation of Korea (NRF) grant funded by the Korea government (MSIT) (No. 2018R1A2B6007837) and Creative-Pioneering Researchers Program through Seoul National University (SNU).


\clearpage 
\widetext
\setcounter{section}{0}
\setcounter{equation}{0}
\setcounter{figure}{0} 
\setcounter{table}{0} 
\renewcommand{\theequation}{S.\arabic{equation}}
\renewcommand\thefigure{S\arabic{figure}} 
\setcounter{page}{1}

\large
\begin{center}
{\bf Supplemental Material:
Tunable quantum interference effect on magnetoconductivity in few-layer black phosphorus}
\end{center}
\normalsize

\section{Eigenstates of few-layer BP}
\label{sec:H parametrization}
In this section, we present the eigenstates for the various phases of few-layer BP, following the parametrization of Park {\em et al.} \cite{SM_Park2019}. For the insulator phase and SDTP \((E_{\rm{g}}\geq0)\), we can rewrite the Hamiltonian in Eq.~(\ref{eq:H}) as
\begin{eqnarray}
\label{eq:SM_Hamiltonian}
H=\varepsilon_{0}(\cos{\phi_{\bm{k}}}\sigma_{x}+\sin{\phi_{\bm{k}}}\sigma_{y}),
\end{eqnarray}
where \(\varepsilon_{0}\equiv \sqrt{\left(\frac{\hbar^{2}k^{2}_{x}}{2m^{*}}+\frac{E_{\rm{g}}}{2}\right)^{2}+ (\hbar v_{y}k_{y})^{2}}\), and thus the energy eigenvalues are given by \(E=\pm \varepsilon_{0}\). The angle \(\phi_{\bm{k}}\) is defined via \(\varepsilon_{0} \cos{\phi_{\bm{k}}}=\frac{\hbar^{2}k^{2}_{x}}{2m^{*}}+\frac{E_{\rm{g}}}{2}\) and \(\varepsilon_{0} \sin{\phi_{\bm{k}}}=\hbar v_{y}k_{y}\). For the upper band \((E=+\varepsilon_{0})\), the eigenstate corresponding to \(\bm{k}\) reads 
\(\ket{\bm{k}}=\frac{1}{\sqrt2}(1, e^{i\phi_{\bm{k}}})^{T}\). Note that \(\phi_{-\bm{k}}=-\phi_{\bm{k}}\) for the insulator phase and SDTP. 

For the DSM phase with a sufficiently low Fermi energy \((E_{\rm{F}}\ll E_{\rm{g}})\), the Hamiltonian near the positive/negative node \(K^{\pm}\) can be rewritten as 
\begin{eqnarray}
H=\hbar v_{0}\kappa(\pm\cos{\phi_{\bm{\kappa}}}\sigma_{x}+\sin{\phi_{\bm{\kappa}}}\sigma_{y}),
\end{eqnarray}
where we adopt the parametrization \(v_{0}\kappa\cos{\phi_{\bm{\kappa}}}=v_{x}\left(k_{x}-K^{\pm}_{x}\right)\) and \(v_{0}\kappa\sin{\phi_{\bm{\kappa}}}=v_{y}k_{y}\). Note that we introduce an effective momentum \(\bm{\kappa}\) to describe momentum states near the node. The corresponding eigenstate for the upper band is given by \(\ket{\bm{\kappa}}=\frac{1}{\sqrt2}(1, \pm e^{\pm i\phi_{\bm{\kappa}}})^{T}\), and thus we have \(\phi_{-\bm{\kappa}}=\phi_{\bm{\kappa}}+\pi\). 

\section{Ladder vertex corrections}
\label{sec:ladder vertex corrections}

In this section, we provide a brief explanation on the ladder vertex correction for elastic impurity scattering. In a many-body diagrammatic approach, the dc conductivity can be obtained from the current-current correlation function. First, let us consider a \(d-\)dimensional isotropic system. The single bubble diagram captures the Drude conductivity \(\sigma^{\rm{Drude}}=e^{2}N_{0}\mathcal{D}\) where \(\mathcal{D}=v_{\rm{F}}^{2}\tau^{\rm{qp}}/d\) is the diffusion constant with the quasiparticle lifetime \(\tau^{\rm{qp}}\).
The ladder vertex correction is the leading impurity correction to the current vertex, which obeys the self-consistent Dyson's equation \cite{SM_Mahan, SM_Coleman} represented in Figs.~\ref{fig:2}(a) and ~\ref{fig:2}(b). In a single-band isotropic system, it is equivalent to replacing \(\tau^{\rm{qp}}\) in \(\mathcal{D}\) by the transport relaxation time \(\tau^{\rm{tr}}\) which satisfies \cite{SM_Ashcroft}
\begin{equation}
\label{eq:relaxation_time_isotropic}
{1\over \tau_{\bm k}^{\rm tr}}=\int {d^d k' \over (2\pi)^d} W_{\bm{k}',\bm{k}} (1-\cos\theta_{\bm{k}\bm{k}'}).
\end{equation}


In an anisotropic system, the relation in Eq.~(\ref{eq:relaxation_time_isotropic}) is no longer valid, and a further systematic relation is required fully considering the anisotropy of the system.
In a diagrammatic approach, the ladder vertex can be related to the transport relaxation time by \(\tau_{\bm{k}}^{(i)}= \tau_{\bm{k}}^{\rm qp} \Lambda^{(i){\rm{AR}}} (\bm{k},\xi_{\bm{k}},\xi_{\bm{k}})\), where the ladder vertex \(\Lambda^{(i){\rm{AR}}} (\bm{k},\xi_{\bm{k}},\xi_{\bm{k}})\) modifies the velocity operator along the \(i\)th direction by \(\tilde{v}^{(i)}_{\bm{k}}=v^{(i)}_{\bm{k}} \Lambda^{(i){\rm{AR}}} (\bm{k},\xi_{\bm{k}},\xi_{\bm{k}})=v^{(i)}_{\bm{k}}\tau^{(i)}_{\bm{k}}
/\tau^{\rm{qp}}_{\bm{k}}\) \cite{SM_Kim}. As a result, the Dyson's equation turns out to be equivalent to the integral equation relating transport relaxation times at different states by
\begin{equation}
    1=\int \frac{d^{d}k'}{(2\pi)^d}W_{\bm{k}',\bm{k}}\left(\tau^{(i)}_{\bm{k}}-\frac{v^{(i)}_{\bm{k'}}}{v^{(i)}_{\bm{k}}}\tau^{(i)}_{\bm{k'}}\right),
\end{equation}
which is the generalization of Eq.~(\ref{eq:relaxation_time_isotropic}). Note that for elastic impurity scattering, the transition rate is given by \(W_{\bm{k}',\bm{k}}=\frac{2\pi n_{\rm{imp}}}{\hbar}|V_{\bm{k},\bm{k'}}|^{2}\delta(\xi_{\bm{k}}-\xi_{\bm{k'}})\). Consequently, the dc conductivity incorporating the ladder vertex correction reads
\begin{eqnarray}
\label{eq:dc conductivity}
\sigma_{ij}^{\rm{B}}=ge^{2} \int \frac{d^{d}k}{(2\pi)^d} S^0(\xi_{\bm{k}}) v_{\bm{k}}^{(i)} \tilde{v}_{\bm{k}}^{(j)} \tau_{\bm{k}}^{\rm qp}=ge^{2} \int \frac{d^{d}k}{(2\pi)^d} S^0(\xi_{\bm{k}}) v_{\bm{k}}^{(i)} v_{\bm{k}}^{(j)} \tau_{\bm{k}}^{(j)},
\end{eqnarray}
where the superscript B denotes the Boltzmann conductivity and \(S^{0}(\xi)=-\frac{\partial f^{0}(\xi)}{\partial \xi}=\beta f^{0}(\xi)\left[1-f^{0}(\xi)\right]\). Here, \(f^{0}(\xi)=\left[e^{\beta \xi}+1\right]^{-1}\) is the Fermi-Dirac distribution function in equilibrium with $\beta={1\over k_{\rm B}T}$. In the zero-temperature limit, \(S^{0}(\xi)\) is given by the Dirac-delta function \(\delta(\xi)\).

\section{Cooperon}
\label{sec:cooperon}

\subsection{Bethe-Salpeter equation}
\label{subsec:bethe-salpeter}
This section is devoted to transforming the Bethe-Salpeter equation into a form capturing the full anisotropy of the system. Henceforth, we omit \(\hbar\) for simplicity. The Cooperon obeys the following Bethe-Salpeter equation:
\begin{eqnarray}
\label{eq:bethe-salpeter1}
C^{\rm{AR}}_{\bm{Q}}(\bm{k},\bm{k}')
&=&n_{\rm{imp}}V_{\bm{k'},\bm{k}}V_{\bm{Q}-\bm{k'},\bm{Q}-\bm{k}}+\frac{n_{\rm{imp}}}{\mathcal{V}}\sum_{\bm{p}}V_{\bm{p},\bm{k}}V_{\bm{Q}-\bm{p},\bm{Q}-\bm{k}}C^{\rm{AR}}_{\bm{Q}}(\bm{p},\bm{k}',0,0)G^{\rm{A}}(\bm{p},0)G^{\rm{R}}(\bm{Q}-\bm{p},0) \nonumber \\
&&\approx n_{\rm{imp}}V_{\bm{k'},\bm{k}}V_{-\bm{k'},-\bm{k}}
+\frac{n_{\rm{imp}}}{\mathcal{V}}\sum_{\bm{p}}V_{\bm{p},\bm{k}}V_{\bm{-p},\bm{-k}}C^{\rm{AR}}_{\bm{Q}}(\bm{p},\bm{k}',0,0)\delta(\xi_{\bm{p}})\int d{\xi_{\bm{p}}}G^{\rm{A}}(\bm{p},0)G^{\rm{R}}(\bm{Q}-\bm{p},0),
\end{eqnarray}
where \(V_{\bm{k'},\bm{k}}=\matrixel{\bm{k}'}{V}{\bm{k}}\) is the matrix element of the scattering potential \(V\). 
Here, we assume that \(V_{\bm{Q}-\bm{k}',\bm{Q}-\bm{k}}\approx V_{-\bm{k}',-\bm{k}}\) since the Cooperon diverges as $\bm{Q}\rightarrow 0$ and the dominant contribution comes from small $\bm{Q}$. We compute the integral over energy with the aid of a contour integral method: 
\begin{eqnarray}
\label{eq:two Green's functions contour integral}
  \int d{\xi_{\bm{p}}}G^{\rm{A}}(\bm{p},0)G^{\rm{R}}(\bm{Q}-\bm{p},0)&=& \int d\xi_{\bm{p}}
  \frac{1}{\xi_{\bm{p}}+\frac{i}{2\tau^{\rm{qp}}_{\bm{p}}}}\frac{1}{\xi_{\bm{Q}-\bm{p}}-\frac{i}{2\tau^{\rm{qp}}_{\bm{Q}-\bm{p}}}}
  \nonumber\\&=&\frac{2\pi i}{\bm{Q}\cdot \bm{v}_{\bm{p}}+\frac{i}{2}\left(\frac{1}{\tau^{\rm{qp}}_{\bm{p}}}+\frac{1}{\tau^{\rm{qp}}_{\bm{p}-\bm{Q}}}\right)},
\end{eqnarray}
where we used \(\xi_{\bm{Q}-\bm{p}}=\xi_{\bm{p}-\bm{Q}}\approx \xi_{\bm{p}}-\bm{Q}\cdot \frac{\partial\xi_{\bm{p}}}{\partial\bm{p}}=\xi_{\bm{p}}-\bm{Q}\cdot \bm{v}_{\bm{p}}\). We deal with the denominator of the above equation using the relation between the self-energy and quasiparticle lifetime, \(\Sigma(\bm{p})=\lim_{\omega \rightarrow 0}\Sigma(\bm{p},i\omega_n \rightarrow \omega+i0^{+})= \frac{-i}{2\tau^{\rm{qp}}_{\bm{p}}}\). Then 
\begin{eqnarray}
\frac{i}{2\tau^{\rm{qp}}_{\bm{p}-\bm{Q}}}=\frac{i}{2\tau^{\rm{qp}}_{\bm{p}}}+\frac{i}{2}\frac{\partial}{\partial\bm{p}}\left(\frac{1}{\tau^{\rm{qp}}_{\bm{p}}}\right)\cdot (-\bm{Q})=\frac{i}{2\tau^{\rm{qp}}_{\bm{p}}}+\bm{Q}\cdot \frac{\partial \Sigma(\bm{p})}{\partial \bm{p}}.
\end{eqnarray}
Using the Ward identity, \(v_{\bm{k}}^{(j)}\Lambda^{(j)}(\bm{k},i\omega_n,i\omega_n)=v^{(j)}_{\bm{k}}+\frac{\partial\Sigma(\bm{k},i\omega_n)}{\partial k^{(j)}}\) in anisotropic systems \cite{SM_Kim}, we have
\begin{eqnarray}
\bm{Q}\cdot \bm{v}_{\bm{p}}+\frac{i}{2}\left(\frac{1}{\tau^{\rm{qp}}_{\bm{p}}}+\frac{1}{\tau^{\rm{qp}}_{\bm{p}-\bm{Q}}}\right)=\frac{1}{\tau^{\rm{qp}}_{\bm{p}}}\big(i+\sum_{j}Q^{(j)}v^{(j)}_{\bm{p}}\tau^{(j)}_{\bm{p}}\big).
\end{eqnarray}
Accordingly, the Bethe-Salpeter equation can be rewritten as 
 \begin{eqnarray}
 \label{eq:bethe-salpeter2}
 C^{\rm{AR}}_{\bm{Q}}(\bm{k},\bm{k}')&=&n_{\rm{imp}}V_{\bm{k}',\bm{k}}V_{-\bm{k}',-\bm{k}}+\frac{2\pi n_{\rm{imp}}}{\mathcal{V}}\sum_{\bm{p}}\delta(\xi_{\bm{p}})\frac{V_{\bm{p},\bm{k}}V_{-\bm{p},-\bm{k}}C^{\rm{AR}}_{\bm{Q}}(\bm{p},\bm{k}')\tau^{\rm{qp}}_{\bm{p}}}{1-if_{\bm{Q}}(\bm{p})}\nonumber\\
 &\approx& n_{\rm{imp}}V_{\bm{k'},\bm{k}}V_{-\bm{k'},-\bm{k}}+\frac{2\pi n_{\rm{imp}}}{\mathcal{V}}\sum_{\bm{p}}\delta(\xi_{\bm{p}})V_{\bm{p},\bm{k}}V_{-\bm{p},-\bm{k}}C^{\rm{AR}}_{\bm{Q}}(\bm{p},\bm{k}')\tau^{\rm{qp}}_{\bm{p}}\left[1+if_{\bm{Q}}(\bm{p})-f_{\bm{Q}}^{2}(\bm{p})\right],
 \end{eqnarray}
where \(f_{\bm{Q}}(\bm{p})\equiv \sum_{j}Q^{(j)}v^{(j)}_{\bm{p}}\tau^{(j)}_{\bm{p}}\). Note that terms of order higher than \(Q^{2}\) are ignored. 

\subsection{Cooperon ansatz}
\label{subsec:ansatz}

Now, we are in a position to solve the Bethe-Salpeter equation in Eq.~(\ref{eq:bethe-salpeter2}). We note that there appears \(V_{\bm{k'},\bm{k}}V_{-\bm{k'},-\bm{k}}\) in Eq.~(\ref{eq:bethe-salpeter2}). Let us assume that \(V_{-\bm{k}',-\bm{k}}\equiv V^{*}_{\bm{k}',\bm{k}} F(\bm{k},\bm{k}')\), where the phase factor \(F(\bm{k},\bm{k}')\) is determined from the electronic structure of the system. Here, we consider systems where \(F(\bm{k}_{1},\bm{k}_{3})=F(\bm{k}_{1}
,\bm{k}_{2})F(\bm{k}_{2},\bm{k}_{3})\) holds, such as in few-layer BP. Introducing \(\tilde{C}^{\rm{AR}}_{\bm{Q}}(\bm{k},\bm{k}')
\equiv C^{\rm{AR}}_{\bm{Q}}(\bm{k},\bm{k}')F(\bm{k}',\bm{k})\), the Bethe-Salpeter equation reduces to 

\begin{eqnarray}
\label{eq:SM_bethe-salpeter}
\tilde{C}^{\rm{AR}}_{\bm{Q}}(\bm{k},\bm{k}')
&=& n_{\rm{imp}}|V_{\bm{k'},\bm{k}}|^{2}+\frac{2\pi n_{\rm{imp}}}{\mathcal{V}}\sum_{\bm{p}}\delta(\xi_{\bm{p}})|V_{\bm{p},\bm{k}}|^{2}\tilde{C}^{\rm{AR}}_{\bm{Q}}(\bm{p},\bm{k}')\tau^{\rm{qp}}_{\bm{p}}\left[1+if_{\bm{Q}}(\bm{p})-f_{\bm{Q}}^{2}(\bm{p})\right].
\end{eqnarray}
Performing \(\frac{1}{\mathcal{V}}\sum_{\bm{k}}\delta(\xi_{\bm{k}})\) on both sides of Eq.~(\ref{eq:SM_bethe-salpeter}), we have
\begin{eqnarray}
\label{eq:SM_bethe-salpeter2}
(2\pi\tau^{\rm{qp}}_{\bm{k'}})^{-1}&=&\frac{1}{\mathcal{V}}\sum_{\bm{p}}\delta(\xi_{\bm{p}})\tilde{C}^{\rm{AR}}_{\bm{Q}}(\bm{p},\bm{k}')\left[-if_{\bm{Q}}(\bm{p})+f_{\bm{Q}}^{2}(\bm{p})\right],
\end{eqnarray}
where we use the definition of the quasiparticle lifetime
\(\frac{1}{\tau^{\rm{qp}}_{\bm{p}}}=\frac{2\pi n_{\rm{imp}}}{\mathcal{V}}\sum_{\bm{k}}|V_{\bm{p},\bm{k}}|^{2}\delta(\xi_{\bm{k}})\). We consider the following ansatz for Eq.~(\ref{eq:SM_bethe-salpeter2}):
\begin{eqnarray}
\label{eq:ansatz_insulator}
\tilde{C}^{\rm{AR}}_{\bm{Q}}(\bm{k},\bm{k'})=\frac{(2\pi N_{0}\tau^{\rm{qp}}_{\bm{k}}\tau^{\rm{qp}}_{\bm{k'}})^{-1}}{\sum_{i,j}D_{ij}Q_{i}Q_{j}},
\end{eqnarray}
where \(D_{ij}\) denotes the diffusion coefficients. Plugging this ansatz into Eq.~(\ref{eq:SM_bethe-salpeter2}), we have
\begin{eqnarray}
\label{diffusion tensor_insulator}
D_{ij}=\frac{1}{N_{0}\mathcal{V}}\sum_{\bm{p}}\delta(\xi_{\bm{p}})v_{\bm{p}}^{(i)}v_{\bm{p}}^{(j)}\tau^{(i)}_{\bm{p}}\tau^{(j)}_{\bm{p}}(\tau^{\rm{qp}}_{\bm{p}})^{-1}=\frac{1}{N_{0}\mathcal{V}}\sum_{\bm{p}}\delta(\xi_{\bm{p}})(v_{\bm{p}}^{(i)})^{2}(\tau^{(i)}_{\bm{p}})^{2}(\tau^{\rm{qp}}_{\bm{p}})^{-1}\delta_{ij}.
\end{eqnarray}
Since the most divergent terms in the left and right-hand sides are identical as we plug this ansatz into Eq.~(\ref{eq:SM_bethe-salpeter}), it is consistent with the ansatz obtained from the iterative method in previous works \cite{SM_Suzuura, SM_Lu, SM_Chen}. Finally, we obtain the general ansatz for the Bethe-Salpeter equation as
\begin{eqnarray}
C^{\rm{AR}}_{\bm{Q}}(\bm{k},\bm{k}')=\frac{(2\pi N_{0}\tau^{\rm{qp}}_{\bm{k}}\tau^{\rm{qp}}_{\bm{k}'})^{-1}}{\sum_{i,j}D_{ij}Q_{i}Q_{j}}F(\bm{k},\bm{k}').
\end{eqnarray}

\section{WL and WAL corrections}

\subsection{Insulator phase and SDTP}

The quantum correction to the dc conductivity is contributed by a bare Hikami box and two dressed Hikami boxes as illustrated in Fig.~\ref{fig:2}(c). Before computing each term, we note the following identities:
\begin{subequations}
\label{eq:useful formulas}
\begin{eqnarray}
\int d\xi_{\bm{k}}|G^{\rm{R}}(\bm{k},0)|^{4}&=&4\pi (\tau^{\rm{qp}}_{\bm{k}})^{3}, \\
\int d\xi_{\bm{k}}|G^{\rm{R}}(\bm{k},0)|^{2}G^{\rm{R}}(\bm{k},0)&=&-2\pi i(\tau^{\rm{qp}}_{\bm{k}})^{2}.
\end{eqnarray}
\end{subequations}

First, let us compute the bare Hikami box contribution:
\begin{eqnarray}
 \label{eq:insulator_bare_hikami}
    \Delta \sigma^{\rm{bare}}_{ii}&=&\frac{g_{s}e^2}{2\pi\mathcal{V}^2}\sum_{\bm{k},\bm{Q}}|G^{\rm{R}}(\bm{k},0)|^{2}|G^{\rm{R}}(\bm{Q}-\bm{k},0)|^{2}C^{\rm{AR}}_{\bm{Q}}(\bm{k},\bm{Q}-\bm{k})\tilde{v}^{(i)}_{\bm{k}}\tilde{v}^{(i)}_{\bm{Q}-\bm{k}}
     \nonumber\\
     &\approx& \frac{g_{s}e^2}{2\pi\mathcal{V}^2}\sum_{\bm{k},\bm{Q}}|G^{\rm{R}}(\bm{k},0)|^{4}C^{\rm{AR}}_{\bm{Q}}(\bm{k},-\bm{k})\tilde{v}^{(i)}_{\bm{k}}\tilde{v}^{(i)}_{-\bm{k}}
     \nonumber\\
     &=&-\frac{g_{s}e^2}{2\pi}\int \frac{{d^{d}}k}{(2\pi)^d} \left(v^{(i)}_{\bm{k}}\frac{\tau^{(i)}_{\bm{k}}}{\tau^{\rm{qp}}_{\bm{k}}}\right)^{2}\delta(\xi_{\bm{k}})\int d\xi_{\bm{k}}|G^{\rm{R}}(\bm{k},0)|^{4} \int \frac{{d^{d}}Q}{(2\pi)^d}C^{\rm{AR}}_{\bm{Q}}(\bm{k},-\bm{k})\nonumber\\
     &=&-\frac{g_{s}e^2}{\pi N_{0}}\int \frac{{d^{d}}k}{(2\pi)^d}\frac{\left(v^{(i)}_{\bm{k}}\tau^{(i)}_{\bm{k}}\right)^{2}}{\tau^{\rm{qp}}_{\bm{k}}} \delta(\xi_{\bm{k}}) \int \frac{d^d Q}{(2\pi)^d}\frac{1}{\sum_{j}D_{jj}Q^{2}_{j}},
\end{eqnarray}
where \(g_{s}=2\) is the spin degeneracy factor.

Next, let us compute the dressed Hikami box contribution. It includes two diagrams which contribute equally to the dc conductivity. One of them can be computed as follows:
\begin{eqnarray}
 \label{eq:insulator_dressed_hikami}
 \Delta \sigma^{\rm{dressed}}_{ii}&&=\frac{g_{s}e^2}{2\pi\mathcal{V}^3}\sum_{\bm{k},\bm{p},\bm{Q}}n_{\rm{imp}}|V_{\bm{p},\bm{k}}|^{2}|G^{\rm{R}}(\bm{k},0)|^{2}|G^{\rm{R}}(\bm{Q}-\bm{p},0)|^{2}G^{\rm{R}}(\bm{p},0)G^{\rm{R}}(\bm{Q}-\bm{k},0)C^{\rm{AR}}_{\bm{Q}}(\bm{p},\bm{Q}-\bm{k})\tilde{v}^{(i)}_{\bm{k}}\tilde{v}^{(i)}_{\bm{Q}-\bm{p}}
     \nonumber\\
     &&\approx \frac{g_{s}e^2}{2\pi\mathcal{V}^3}\sum_{\bm{k},\bm{p},\bm{Q}}n_{\rm{imp}}|V_{\bm{p},\bm{k}}|^{2}|G^{\rm{R}}(\bm{k},0)|^{2}G^{\rm{R}}(\bm{k},0)|G^{\rm{R}}(\bm{p},0)|^{2}G^{\rm{R}}(\bm{p},0)C^{\rm{AR}}_{\bm{Q}}(\bm{p},-\bm{k})\tilde{v}^{(i)}_{\bm{k}}\tilde{v}^{(i)}_{-\bm{p}}
     \nonumber\\&&
     =-\frac{g_{s}e^2}{2\pi}\int \int \frac{d^d k}{(2\pi)^d} \frac{d^d p}{(2\pi)^d}n_{\rm{imp}}|V_{\bm{p},\bm{k}}|^{2}\left(v^{(i)}_{\bm{k}}\frac{\tau^{(i)}_{\bm{k}}}{\tau^{\rm{qp}}_{\bm{k}}}\right)\left(v^{(i)}_{\bm{p}}\frac{\tau^{(i)}_{\bm{p}}}{\tau^{\rm{qp}}_{\bm{p}}}\right)\delta(\xi_{\bm{k}}) \delta(\xi_{\bm{p}})
     \nonumber\\&& \times \int d\xi_{\bm{k}}|G^{\rm{R}}(\bm{k},0)|^{2}G^{\rm{R}}(\bm{k},0) \int d\xi_{\bm{p}}|G^{\rm{R}}(\bm{p},0)|^{2}G^{\rm{R}}(\bm{p},0) \int \frac{d^d Q}{(2\pi)^d}C^{\rm{AR}}_{\bm{Q}}(\bm{p},-\bm{k})
     \nonumber\\
     &&= \frac{g_{s}e^2}{N_{0}} \int \int \frac{d^d k}{(2\pi)^d}\frac{d^d p}{(2\pi)^d} n_{\rm{imp}}|V_{\bm{p},\bm{k}}|^{2} v^{(i)}_{\bm{k}}\tau^{(i)}_{\bm{k}}v^{(i)}_{\bm{p}}\tau^{(i)}_{\bm{p}}\delta(\xi_{\bm{k}})\delta(\xi_{\bm{p}})\int \frac{d^d Q}{(2\pi)^d}\frac{1}{\sum_{j}D_{jj}Q^{2}_{j}}
     \nonumber\\&&=\frac{g_{s}e^2}{2\pi N_{0}}\int \frac{d^d p}{(2\pi)^d}\left(v^{(i)}_{\bm{p}}\frac{\tau^{(i)}_{\bm{p}}}{\tau^{\rm{qp}}_{\bm{p}}}-v^{(i)}_{\bm{p}}\right)v^{(i)}_{\bm{p}}\tau^{(i)}_{\bm{p}}\delta(\xi_{\bm{p}})\int \frac{d^d Q}{(2\pi)^d}\frac{1}{\sum_{j}D_{jj}Q^{2}_{j}}
     \nonumber\\&&=-\frac{1}{2}\Delta \sigma^{\rm{bare}}_{ii}-\frac{g_{s}e^2}{2\pi N_{0}}\int \frac{d^d p}{(2\pi)^d}(v^{(i)}_{\bm{p}})^{2}\tau^{(i)}_{\bm{p}}\delta(\xi_{\bm{p}})\int \frac{d^d Q}{(2\pi)^d}\frac{1}{\sum_{j}D_{jj}Q^{2}_{j}}.
 \end{eqnarray}
Therefore, the total quantum correction to the dc conductivity is 
\begin{eqnarray}
\label{eq:insulator_hikami_total1}
\Delta \sigma_{ii}&=&\Delta \sigma^{\rm{bare}}_{ii}+2\Delta \sigma^{\rm{dressed}}_{ii} \nonumber\\
&=&-\frac{g_{s}e^2}{\pi N_{0}}\int \frac{d^d p}{(2\pi)^d}(v^{(i)}_{\bm{p}})^{2}\tau^{(i)}_{\bm{p}}\delta(\xi_{\bm{p}})\int \frac{d^d Q}{(2\pi)^d}\frac{1}{\sum_{j}D_{jj}Q^{2}_{j}},
\end{eqnarray}
which indicates that the insulator phase and SDTP exhibit the WL effect. To compute the \(Q-\)integral, let us parametrize the momentum as \(\tilde{Q}_{x}=\sqrt{\frac{D_{xx}}{D}}Q_{x}\), \(\tilde{Q}_{y}=\sqrt{\frac{D_{yy}}{D}}Q_{y}\), and \(\tilde{Q}^{2}=\tilde{Q}^{2}_{x}+\tilde{Q}^{2}_{y}\) with \(D\equiv \sqrt{D_{xx}D_{yy}}\). Accordingly, the \(Q-\)integral can be rewritten as \begin{eqnarray}
\int \frac{d \tilde{Q}_{x} d \tilde{Q}_{y}}{(2\pi)^2}\frac{1}{D(\tilde{Q}^{2}_{x}+\tilde{Q}^{2}_{y})}=\int_{\ell^{-1}_{\phi}}^{\ell^{-1}_{e}} \frac{d\tilde{Q}}{(2\pi)}\frac{1}{D\tilde{Q}}=\frac{1}{2\pi D}\ln{\left(\frac{\ell_{\phi}}{\ell_{e}}\right)},
\end{eqnarray}
where \(\ell_{\phi}\) is the phase coherence length and \(\ell_{e}\) is the mean-free path. Here, we assume that the lower and upper cutoffs of the integral are given by \(\ell^{-1}_{\phi}\) and \(\ell^{-1}_{e}\), respectively. Restoring \(\hbar\), we finally obtain the WL correction as
\begin{eqnarray}
\label{eq:insulator_hikami_total2}
\Delta \sigma_{ii}=-\frac{g_{s}e^2}{2\pi^2 N_{0}D\hbar}\ln{\left(\frac{\ell_{\phi}}{\ell_{e}}\right)}\int \frac{d^d p}{(2\pi)^d}(v^{(i)}_{\bm{p}})^{2}\tau^{(i)}_{\bm{p}}\delta(\xi_{\bm{p}})=-\frac{1}{2\pi^2 N_{0}D\hbar}\ln{\left(\frac{\ell_{\phi}}{\ell_{e}}\right)}\sigma^{\rm{B}}_{ii},
\end{eqnarray}
which applies not only to few-layer BP but also to a general 2D anisotropic system. Importantly, the quantum correction is proportional to the Boltzmann conductivity, and thus the ratio of the quantum correction to the Boltzmann conductivity does not depend on the direction regardless of the anisotropy of the system. Rewriting Eq.~(\ref{eq:insulator_hikami_total1}) as 
\begin{eqnarray}
\frac{\Delta \sigma_{ii}}{\sigma_{ii}^{\rm B}}=-\frac{1}{\pi N_{0}}\int \frac{d^d Q}{(2\pi)^d}\frac{1}{\sum_{j}D_{jj}Q^{2}_{j}},
\end{eqnarray}
we note that this direction-independent ratio generally holds even in 3D systems. 

\subsection{DSM phase with intranode scattering}
Before studying the quantum interference effect in the DSM phase due to intranode scattering, we note several identities: \(N_{0}=\frac{g_{s}E_{\rm{F}}}{\pi v_{x}v_{y}}\), \((\tau^{\rm{qp}})^{-1}=\frac{n_{\rm{imp}}V^{2}_{\rm{intra}}E_{\rm{F}}}{2v_{x}v_{y}}\), \((\tau^{\rm{tr}})^{-1}=\frac{n_{\rm{imp}}V^{2}_{\rm{intra}}E_{\rm{F}}}{4v_{x}v_{y}}\), \(D_{xx}=\frac{2E_{\rm{F}}\tau^{\rm{qp}}}{N_{0}\pi}\left(\frac{v_{x}}{v_{y}}\right)\), and \(D_{yy}=\frac{2E_{\rm{F}}\tau^{\rm{qp}}}{N_{0}\pi}\left(\frac{v_{y}}{v_{x}}\right)\). Note that the transport relaxation time \(\tau^{\rm{tr}}\) is isotropic and equals \(2\tau^{\rm{qp}}\). Using these identities, we obtain the explicit form of the Cooperon as
\begin{eqnarray}
C^{\rm{AR}}_{\bm{Q}}(\bm{\kappa},\bm{\kappa}')=\frac{1}{4(\tau^{\rm{qp}})^{2}}\frac{n_{\rm{imp}}V^{2}_{\rm{intra}}e^{i(\phi-\phi')}}{v_{x}^{2}Q^{2}_{x}+v_{y}^{2}Q^{2}_{y}}.
\end{eqnarray}
We note that for backscattering, the phase factor becomes \(e^{i(\phi-\phi')}=-1\), and thus the Cooperon has a negative value. Therefore, intranode scattering in the DSM phase induces the WAL effect in the \(E_{\rm{F}}\ll |E_{\rm{g}}|\) limit, as will be demonstrated in the following. 

We are now in a position to compute the quantum correction of the DSM phase in the presence of intranode scattering. First, let us compute \(\Delta \sigma^{\rm{bare}}_{xx}\):
\begin{eqnarray}
\label{eq:dirac_bare_hikami}
    \Delta \sigma^{\rm{bare}}_{xx}
    &&=\frac{g_{s}e^2}{\pi N_{0}}\int \frac{d^d k}{(2\pi)^d}\frac{\left(v^{(x)}_{\bm{k}}\tau^{(x)}_{\bm{k}}\right)^{2}}{\tau^{\rm{qp}}_{\bm{k}}} \delta(\xi_{\bm{k}}) \int \frac{d^d Q}{(2\pi)^d}\frac{1}{\sum_{j}D_{jj}Q^{2}_{j}}
    \nonumber\\&&=\frac{2g_{s}e^2}{\pi^{2}N_{0}}\left(\frac{v_{x}}{v_{y}}\right)E_{\rm{F}}\tau^{\rm{qp}}\int \frac{d^d Q}{(2\pi)^d}\frac{1}{\sum_{j}D_{jj}Q^{2}_{j}}.
\end{eqnarray}
Next, one of the dressed Hikami boxes reads
\begin{eqnarray}
 \label{eq:dirac_dressed_hikami}
 \Delta \sigma^{\rm{dressed}}_{xx}&&=-\frac{g_{s}e^2}{2\pi}\int \int \frac{d^d k}{(2\pi)^d} \frac{d^d k'}{(2\pi)^d}n_{\rm{imp}}|V_{\bm{k}',\bm{k}}|^{2}F(\bm{k},\bm{k}')\left(v^{(x)}_{\bm{k}}\frac{\tau^{(x)}_{\bm{k}}}{\tau^{\rm{qp}}_{\bm{k}}}\right)\left(v^{(x)}_{\bm{k}'}\frac{\tau^{(x)}_{\bm{k}'}}{\tau^{\rm{qp}}_{\bm{k}'}}\right)\delta(\xi_{\bm{k}}) \delta(\xi_{\bm{k}'})
     \nonumber\\&& \times \int d\xi_{\bm{k}}|G^{\rm{R}}(\bm{k},0)|^{2}G^{\rm{R}}(\bm{k},0) \int d\xi_{\bm{k}'}|G^{\rm{R}}(\bm{k}',0)|^{2}G^{\rm{R}}(\bm{k}',0) \int \frac{d^d Q}{(2\pi)^d}C^{\rm{AR}}_{\bm{Q}}(\bm{k}',-\bm{k})
     \nonumber\\&& = -\frac{g_{s}e^2}{2\pi^{2}N_{0}}\left(\frac{v_{x}}{v_{y}}\right)E_{\rm{F}}\tau^{\rm{qp}}\int \frac{d^d Q}{(2\pi)^d}\frac{1}{\sum_{j}D_{jj}Q^{2}_{j}}=-\frac{1}{4}\Delta \sigma^{\rm{bare}}_{xx}.
\end{eqnarray}

Accordingly, the total quantum correction to the dc conductivity is
\begin{eqnarray}
\label{eq:dirac_hikami_total1}
\Delta \sigma_{xx}&&=\Delta \sigma^{\rm{bare}}_{xx}+2\Delta \sigma^{\rm{dressed}}_{xx}=\frac{1}{2}\Delta \sigma^{\rm{bare}}_{xx}
\nonumber\\&&=\frac{g_{s}e^2}{\pi^{2}N_{0}}\left(\frac{v_{x}}{v_{y}}\right)E_{\rm{F}}\tau^{\rm{qp}}\int \frac{d^d Q}{(2\pi)^d}\frac{1}{\sum_{j}D_{jj}Q^{2}_{j}}.
\end{eqnarray}
The \(Q-\)integral can be computed by introducing \(\tilde{Q}_{x}=\sqrt{\frac{D_{xx}}{D}}Q_{x}\) and \(\tilde{Q}_{y}=\sqrt{\frac{D_{yy}}{D}}Q_{y}\) with \(D=\sqrt{D_{xx}D_{yy}}\), similarly as we did for the insulator phase and SDTP:
\begin{eqnarray}
\int _{\ell^{-1}_{\phi}}^{\ell^{-1}_{\rm{intra}}} \frac{d\tilde{Q}}{(2\pi)}\frac{1}{d\tilde{Q}}=\frac{1}{2\pi D}\ln\left(\frac{\ell_{\phi}}{\ell_{\rm{intra}}}\right)=\frac{N_{0}}{4E_{\rm{F}}\tau^{\rm{qp}}}\ln\left(\frac{\ell_\phi}{\ell_{\rm{intra}}}\right),
\end{eqnarray}
where \(\ell_{\rm{intra}}\) is the mean-free path corresponding to intranode scattering.  Restoring \(\hbar\), we obtain the total quantum correction to the dc conductivity as
\begin{subequations}
\label{eq:dirac_hikami_x_y}
\begin{eqnarray}
\label{eq:dirac_hikami_total_xx}
\Delta \sigma_{xx}&=&\frac{g_{s}e^2}{4\pi^{2}\hbar}\left(\frac{v_{x}}{v_{y}}\right)\ln\left(\frac{\ell_{\phi}}{\ell_{\rm{intra}}}\right), \\
\label{eq:dirac_hikami_total_yy}
\Delta \sigma_{yy}&=&\frac{g_{s}e^2}{4\pi^{2}\hbar}\left(\frac{v_{y}}{v_{x}}\right)\ln\left(\frac{\ell_{\phi}}{\ell_{\rm{intra}}}\right),
\end{eqnarray}
\end{subequations}
or equivalently
\begin{equation}
\label{eq:ratio_dsm_intranode}
\Delta \sigma_{ii}=\frac{n_{\rm{imp}}V^{2}_{\rm{intra}}}{8\pi \hbar^{2} v_{x}v_{y}} \ln\left(\frac{\ell_{\phi}}{\ell_{\rm{intra}}}\right)\sigma^{\rm{B}}_{ii},
\end{equation}
where \(\sigma^{\rm{B}}_{ii}=\frac{2g_{s}e^{2}v^{2}_{i}\hbar}{\pi n_{\rm{imp}}V^{2}_{\rm{intra}}}\) is the Boltzmann conductivity along the \(i\)th direction. Note that the ratio \(\Delta \sigma_{ii}/\sigma^{\rm{B}}_{ii}\) is the same irrespective of the direction.

\subsection{DSM phase with internode scattering}

Using the eigenstates in the DSM phase (see Sec.~\ref{sec:H parametrization}), we can obtain the matrix elements for internode scattering as
\begin{subequations}
\begin{eqnarray}
V_{\bm{\kappa}',\bm{\kappa}}^{+,-}&=&\frac{V_{\rm{inter}}}{2}[1-e^{-i(\phi+\phi')}], \\
V_{-\bm{\kappa}',-\bm{\kappa}}^{-,+}&=&\frac{V_{\rm{inter}}}{2}[1-e^{i(\phi+\phi')}], 
\end{eqnarray}
\end{subequations}
where \(V_{\rm{inter}}\) is internode scattering amplitude, and the superscripts \(+\) and \(-\) denote the positive and negative nodes, respectively. Thus, \(V_{\bm{\kappa}',\bm{\kappa}}^{+,-}V_{-\bm{\kappa}',-\bm{\kappa}}^{-,+}=|V_{\bm{\kappa}',\bm{\kappa}}^{+,-}|^{2}\) and we have the Cooperon ansatz as
\begin{eqnarray}
\label{eq:ansatz_dirac_SR}
C^{\rm{AR}}_{\bm{Q}}(\bm{\kappa},\bm{\kappa'})=\frac{(2\pi N_{0}\tau^{\rm{qp}}_{\bm{\kappa}}\tau^{\rm{qp}}_{\bm{\kappa'}})^{-1}}{\sum_{i,j}D_{ij}Q_{i}Q_{j}}.
\end{eqnarray}

Repeating the steps in the previous section, we obtain the quantum correction due to internode scattering as
\begin{subequations}
\label{eq:dirac_hikami_SR}
\begin{eqnarray}
 \Delta \sigma_{xx}&=&-\frac{g_{s}e^2}{4\pi^2 \hbar}\left(\frac{v_{x}}{v_{y}}\right)\ln{\left(\frac{\ell_{\phi}}{\ell_{\rm{inter}}}\right)},\\  
 \Delta \sigma_{yy}&=&-\frac{g_{s}e^2}{4\pi^2 \hbar}\left(\frac{v_{y}}{v_{x}}\right)\ln{\left(\frac{\ell_{\phi}}{\ell_{\rm{inter}}}\right)},  
\end{eqnarray}
\end{subequations}
or equivalently
\begin{equation}
\label{eq:ratio_dsm_internode}
\Delta \sigma_{ii}=-\frac{n_{\rm{imp}}V^{2}_{\rm{inter}}}{8\pi \hbar^{2} v_{x}v_{y}} \ln\left(\frac{\ell_{\phi}}{\ell_{\rm{inter}}}\right)\sigma^{\rm{B}}_{ii},
\end{equation}
which differs only by the sign from the quantum correction for intranode scattering in Eq.~(\ref{eq:ratio_dsm_intranode}) with \(\ell_{\rm{intra}}\) \((V_{\rm{intra}})\) replaced by \(\ell_{\rm{inter}}\) \((V_{\rm{inter}})\). Note that as for the intranode scattering, the ratio \(\Delta \sigma_{ii}/\sigma^{\rm{B}}_{ii}\) is the same irrespective of the direction.

In the presence of both intranode and internode scatterings, the quantum interference effect in the DSM phase is determined by the dominant scattering process. Thus, we expect that the WAL (WL) effect will occur when intranode (internode) scattering is dominant.

\section{Magnetoconductivity}

\subsection{Insulator phase}
In the following, we restore \(\hbar\) for clarity. According to J. M. Pereira, Jr {\em et al.} \cite{SM_Pereira}, the Landau levels in small \(B\) have a linear dependence on magnetic field, as in the case of a free electron gas.  Thus, for the insulator phase, we adopt the Hamiltonian of an anisotropic free electron gas as \(H=\frac{p^{2}_{x}}{2m_{x}}+\frac{p^{2}_{y}}{2m_{y}}\) with effective masses \(m_{x
,y}\) along each direction. The corresponding Landau levels are given by \(E_{n}=\frac{\hbar eB}{\sqrt{m_{x}m_{y}}c}\left(n+\frac{1}{2}\right)\). Since the ratio between the diffusion coefficients are given by \(\frac{D_{xx}}{D_{yy}}=\frac{m_{y}}{m_{x}}\), the Landau quantization of momentum reads
\begin{equation}
\label{eq:SM_quantization_insulator}
    D_{xx}Q^{2}_{n,x}+D_{yy}Q^{2}_{n,y}=\frac{D}{\ell^{2}_{B}}\left(n+\frac{1}{2}\right).
\end{equation}
Here, we define the magnetic length \(\ell_{B}=\sqrt{\frac{\hbar c}{4eB}}\). Following the above quantization condition, we modify the \(Q-\)integral in Eq.~(\ref{eq:insulator_hikami_total1}) by
\begin{eqnarray}
\label{eq:quantization_dirac3}
&&\sum_{n}\int \frac{d^{2}Q}{(2\pi)^{2}}\frac{1}{D_{xx}Q^{2}_{x}+D_{yy}Q^{2}_{y}}\delta\left[n+\frac{1}{2}-\frac{\ell_{B}^{2}}{D}(D_{xx}Q^{2}_{x}+D_{yy}Q^{2}_{y})\right]
\nonumber\\&&=\sum_{n}\int \frac{d\tilde{Q}_{x}d\tilde{Q}_{y}}{(2\pi)^{2}D\tilde{Q}^{2}}\delta\left(n+\frac{1}{2}-\ell_{B}^{2}\tilde{Q}^{2}\right)= \frac{1}{4\pi D}\sum_{n_{\rm min}}^{n_{\rm max}} \frac{1}{n+\frac{1}{2}},
\end{eqnarray}
where \(n_{\rm{min}}= (\ell_{B}\ell^{-1}_{\phi})^{2}\) and \(n_{\rm{max}}= (\ell_{B}\ell^{-1}_{e})^{2}\). Thus, the \(Q-\)integral can be rewritten in terms of the digamma function \(\Psi(x)\) as
\begin{eqnarray}
\label{eq:quantization_dirac4}
\frac{1}{4\pi D}\left[\Psi\left(\frac{1}{2}+\frac{\ell^{2}_{B}}{\ell^{2}_{e}}\right)-\Psi\left(\frac{1}{2}+\frac{\ell^{2}_{B}}{\ell^{2}_{\phi}}\right)\right],
\end{eqnarray}
where we used \(\Psi(x+N)-\Psi(x)=\sum_{k=0}^{N-1}\frac{1}{x+k}\). Accordingly, we obtain the magnetoconductivity of the insulator phase as
\begin{eqnarray}
\label{eq:MR_insulator}
\Delta\sigma_{ii}(B)=-\frac{g_{s}e^2}{4\pi^{2}N_{0}D\hbar}\left[\Psi\left(\frac{1}{2}+\frac{\ell^{2}_{B}}{\ell^{2}_{e}}\right)-\Psi\left(\frac{1}{2}+\frac{\ell^{2}_{B}}{\ell^{2}_{\phi}}\right)\right]\int \frac{d^{d}p}{(2\pi)^d}[v^{(i)}_{\bm{p}}]^{2}\tau^{(i)}_{\bm{p}}\delta(\xi_{\bm{p}}).
\end{eqnarray}
Since the digamma function follows the asymptotic form \(\Psi\left(\frac{1}{2}+x\right)\approx \ln{x}+\frac{1}{24x^{2}}+\cdots\) for \(x\rightarrow \infty\), the magnetoconductivity reduces to Eq.~(\ref{eq:insulator_hikami_total2}) in the \(B\rightarrow0\) limit. Thus, the ratio between the magnetoconductivity [Eq.~(\ref{eq:MR_insulator})] and the dc conductivity [Eq.~(\ref{eq:dc conductivity})] reads
\begin{equation}
    \frac{\Delta\sigma_{ii}(B)-\Delta\sigma_{ii}(0)}{\sigma^{\rm{B}}_{ii}}=-\frac{1}{4\pi^{2}N_{0}D\hbar}\left[\Psi\left(\frac{1}{2}+\frac{\ell^{2}_{B}}{\ell^{2}_{e}}\right)-\Psi\left(\frac{1}{2}+\frac{\ell^{2}_{B}}{\ell^{2}_{\phi}}\right)-2\ln{\left(\frac{\ell_{\phi}}{\ell_{e}}\right)}\right].
\end{equation}
Again, the ratio is irrespective of the direction.

\subsection{DSM phase}
Now, let us consider the DSM phase. Applying an external magnetic field perpendicular to the \(xy\) plane, say \(\bm{B}=(0,0,B)\), the crystal momentum is transformed as \(\hbar \bm{k}=\bm{p}+{e\over c}\bm{A}\) where \(\bm{p}\) is the canonical momentum and \(\bm{A}\) is the vector potential. Thus, the Hamiltonian becomes 
\begin{eqnarray}
\label{eq:landau level_dirac2}
H=v_{x}p_{x}\sigma_{x}+v_{y}\left(p_{y}+\frac{eBx}{c}\right)\sigma_{y},
\end{eqnarray}
where we chose the Landau gauge \(\bm{A}=(0,Bx,0)\) for the vector potential. Then we have the \(n\)th Landau level as
\begin{eqnarray}
\label{eq:landau level_dirac3}
E_{n}^{2}=v_{x}^{2}p^{2}_{x}+v_{y}^{2}\left(p_{y}+\frac{eBx}{c}\right)^{2}-\frac{\hbar eBv_{x}v_{y}}{c}=\frac{2\hbar eBv_{x}v_{y}n}{c}.
\end{eqnarray}
Note that the momentum along the \(y\) axis is a good quantum number. Thus, the crystal momentum follows the quantization \(\hbar^{2}v_{x}^{2}k^{2}_{x}+\hbar^{2}v_{y}^{2}k_{y}^{2}=v_{x}^{2}p^{2}_{x}+v_{y}^{2}\left(p_{y}+\frac{eBx}{c}\right)^{2}=\frac{2\hbar eBv_{x}v_{y}}{c}\left(n+\frac{1}{2}\right)\). The quantization for \(\bm{Q}=\bm{k}+\bm{k}'\) is obtained by doubling the magnetic field:
\begin{eqnarray}
\label{quantization_dirac1}
v_{x}^{2}Q^{2}_{n,x}+v_{y}^{2}Q_{n,y}^{2}=\frac{4eB}{\hbar c}v_{x}v_{y}\left(n+\frac{1}{2}\right),
\end{eqnarray}
which is equivalent to Eq.~(\ref{eq:SM_quantization_insulator}) in the insulator phase. Therefore, we compute the magnetoconductivity in the DSM phase in a similar manner as in the insulator phase:
\begin{subequations}
\begin{eqnarray}
\label{eq:MR_dirac_xx}
\Delta\sigma_{xx}(B)&=&\frac{g_{s}e^2}{8\pi^2 \hbar}\left(\frac{v_{x}}{v_{y}}\right)\left[\Psi\left(\frac{1}{2}+\frac{\ell^{2}_{B}}{\ell^{2}_{\rm{intra}}}\right)-\Psi\left(\frac{1}{2}+\frac{\ell^{2}_{B}}{\ell^{2}_{\phi}}\right)\right], \\
\label{eq:MR_dirac_yy}
\Delta\sigma_{yy}(B)&=&\frac{g_{s}e^2}{8\pi^2 \hbar}\left(\frac{v_{y}}{v_{x}}\right)\left[\Psi\left(\frac{1}{2}+\frac{\ell^{2}_{B}}{\ell^{2}_{\rm{intra}}}\right)-\Psi\left(\frac{1}{2}+\frac{\ell^{2}_{B}}{\ell^{2}_{\phi}}\right)\right],
\end{eqnarray}
\end{subequations}
or equivalently
\begin{eqnarray}
\label{eq:41}
\frac{\Delta \sigma_{ii}(B)-\Delta \sigma_{ii}(0)}{\sigma^{\rm{B}}_{ii}}&=&\frac{n_{\rm{imp}}V^{2}_{\rm{intra}}}{16\pi \hbar^{2}v_{x}v_{y}}  \left[\Psi\left(\frac{1}{2}+\frac{\ell^{2}_{B}}{\ell^{2}_{\rm{intra}}}\right)-\Psi\left(\frac{1}{2}+\frac{\ell^{2}_{B}}{\ell^{2}_{\phi}}\right)-2\ln{\left(\frac{\ell_{\phi}}{\ell_{\rm{intra}}}\right)}\right]
\nonumber\\
&=&\frac{\hbar}{8\pi E_{\rm{F}}\tau_{\rm{intra}}^{\rm{qp}}}\left[\Psi\left(\frac{1}{2}+\frac{\ell^{2}_{B}}{\ell^{2}_{\rm{intra}}}\right)-\Psi\left(\frac{1}{2}+\frac{\ell^{2}_{B}}{\ell^{2}_{\phi}}\right)-2\ln{\left(\frac{\ell_{\phi}}{\ell_{\rm{intra}}}\right)}\right].
\end{eqnarray}
We note that in the \(B\rightarrow 0\) limit, Eqs.~(\ref{eq:MR_dirac_xx}) and ~(\ref{eq:MR_dirac_yy}) reduce to Eq.~(\ref{eq:dirac_hikami_total_xx}) and ~(\ref{eq:dirac_hikami_total_yy}), respectively. Similarly, we can compute the magnetoconductivity for internode scattering, which only differs from the result in Eq.~(\ref{eq:41}) by the sign, with \(\ell_{\rm intra}\) (\(\tau^{\rm qp}_{\rm intra}\)) replaced by \(\ell_{\rm inter}\) (\(\tau^{\rm qp}_{\rm inter}\)). 

\subsection{SDTP}

Following Petra Dietl {\em et al.} \cite{SM_Dietl}, we have the Landau quantization for the momentum \(\bm{Q}\) at the SDTP as follows:
\begin{eqnarray}
\frac{\hbar^{4}Q^{4}_{x}}{4m^{*2}}+\hbar^{2}v_{y}^{2}Q_{y}^{2}=A^{2}\left(\frac{\hbar}{2\ell_{B}^{2}}\right)^{\frac{4}{3}}\left(\frac{v_{y}^{2}}{m^{*}}\right)^{\frac{2}{3}}\left(n+\frac{1}{2}\right)^{\frac{4}{3}},
\end{eqnarray}
where \(A\approx1.17325\). Considering the Landau quantization, we rewrite the \(Q-\)integral in Eq.~(\ref{eq:insulator_hikami_total1}) for the quantum correction as
\begin{eqnarray}
\sum_{n}\int \frac{dQ_{x}dQ_{y}}{(2\pi)^{2}}\frac{1}{D_{xx}Q^{2}_{x}+D_{yy}Q^{2}_{y}}\delta\left[n+\frac{1}{2}-\left(\frac{\hbar^{4}Q^{4}_{x}}{4m^{*2}}+\hbar^{2}v_{y}^{2}Q_{y}^{2}\right)^{\frac{3}{4}}\times\frac{\sqrt{m^{*}}}{v_{y}}\frac{2\ell_{B}^{2}}{\hbar^{2}A^{\frac{3}{2}}}\right].
\end{eqnarray}
The momentum \(\tilde{Q}=\sqrt{\tilde{Q}^{2}_{x}+\tilde{Q}^{2}_{y}}\) is bounded as \(\ell_{\phi}^{-1}\leq \tilde{Q} \leq \ell_{e}^{-1}\). We rewrite the \(Q-\)integral as
\begin{eqnarray}
\label{eq:q-integral1}
\sum_{n}\int \frac{d\tilde{Q}_{x}d\tilde{Q}_{y}}{(2\pi)^{2}D}\frac{1}{\tilde{Q}_{x}^{2}+\tilde{Q}_{y}^{2}}\delta\left[n+\frac{1}{2}-\left(\frac{\hbar^{4}\tilde{Q}^{4}_{x}}{4m^{*2}}\left(\frac{D}{D_{xx}}\right)^{2}+\alpha\hbar^{2}v_{y}^{2}\tilde{Q}_{y}^{2}\left(\frac{D}{D_{yy}}\right)\right)^{\frac{3}{4}}\right],
\end{eqnarray}
where \(\alpha\equiv\frac{\sqrt{m^{*}}}{v_{y}}\frac{2\ell_{B}^{2}}{\hbar^{2}A^{\frac{3}{2}}}\). Note that we consider the region where \(\tilde{Q}_{x}\) and \(\tilde{Q}_{y}\) are positive. To calculate the \(Q_{y}-\)integral first, we transform the delta function in Eq.~(\ref{eq:q-integral1}) into 
\begin{eqnarray}
\frac{\delta\left[\tilde{Q}_{y}-\frac{1}{\hbar v_{y}}\sqrt{\frac{D_{yy}}{D}}\sqrt{  \frac{1}{\alpha^{\frac{4}{3}}}\left(n+\frac{1}{2}\right)^{\frac{4}{3}}-\frac{\hbar^{4}\tilde{Q}^{4}_{x}}{4m^{*2}}\left( \frac{D}{D_{xx}}\right)^{2}   }\right]}{\frac{3\alpha}{4}\left[\frac{\hbar^{4}\tilde{Q}^{4}_{x}}{4m^{*2}}\left(\frac{D}{D_{xx}}\right)^{2}+\hbar^{2}v_{y}^{2}\tilde{Q}_{y}^{2}\left(\frac{D}{D_{yy}}\right)\right]^{-\frac{1}{4}}\times\hbar^{2}v^{2}_{y}\frac{D}{D_{yy}}\times 2\tilde{Q}_{y}
}.
\end{eqnarray}

We deal with the above integral differently based on the region \(\tilde{Q}_{x}\) lies in: 1) \(\ell_{\phi}^{-1} \leq|\tilde{Q}_{x}|\leq \ell_{e}^{-1}: 0\leq |\tilde{Q}_{y}|\leq \sqrt{\ell_{e}^{-2}-\tilde{Q}_{x}^{2}}\), 2) \(|\tilde{Q}_{x}|\leq \ell_{\phi}^{-1}: \sqrt{\ell_{\phi}^{-2}-\tilde{Q}_{x}^{2}}\leq |\tilde{Q}_{y}|\leq \sqrt{\ell_{e}^{-2}-\tilde{Q}_{x}^{2}}\). Thus, Eq.~(\ref{eq:q-integral1}) can be rewritten as
\begin{eqnarray}
\label{eq:q-integral2}
&&4\sum_{n}\int_{\ell^{-1}_{\phi}}^{\ell^{-1}_{e}} \frac{d\tilde{Q}_{x}}{(2\pi)^{2}D}\int_{0}^{\sqrt{\ell^{-2}_{e}-\tilde{Q}^{2}_{x}}} \frac{d\tilde{Q}_{y}}{\tilde{Q}_{x}^{2}+\tilde{Q}_{y}^{2}}\times\frac{\delta\left[\tilde{Q}_{y}-\frac{1}{\hbar v_{y}}\sqrt{\frac{D_{yy}}{D}}\sqrt{  \frac{1}{\alpha^{\frac{4}{3}}}\left(n+\frac{1}{2}\right)^{\frac{4}{3}}-\frac{\hbar^{4}\tilde{Q}^{4}_{x}}{4m^{*2}}\left( \frac{D}{D_{xx}}\right)^{2}   }\right]}{\frac{3\alpha}{4}\left[\frac{\hbar^{4}\tilde{Q}^{4}_{x}}{4m^{*2}}\left(\frac{D}{D_{xx}}\right)^{2}+\hbar^{2}v_{y}^{2}\tilde{Q}_{y}^{2}\left(\frac{D}{D_{yy}}\right)\right]^{-\frac{1}{4}}\times\hbar^{2}v^{2}_{y}\frac{D}{D_{yy}}\times 2\tilde{Q}_{y}
}\nonumber\\
&&+4\sum_{n}\int_{0}^{\ell^{-1}_{\phi}} \frac{d\tilde{Q}_{x}}{(2\pi)^{2}D}\int_{\sqrt{\ell^{-2}_{\phi}-\tilde{Q}^{2}_{x}}}^{\sqrt{\ell^{-2}_{e}-\tilde{Q}^{2}_{x}}} \frac{d\tilde{Q}_{y}}{\tilde{Q}_{x}^{2}+\tilde{Q}_{y}^{2}}\times\frac{\delta\left[\tilde{Q}_{y}-\frac{1}{\hbar v_{y}}\sqrt{\frac{D_{yy}}{D}}\sqrt{  \frac{1}{\alpha^{\frac{4}{3}}}\left(n+\frac{1}{2}\right)^{\frac{4}{3}}-\frac{\hbar^{4}\tilde{Q}^{4}_{x}}{4m^{*2}}\left( \frac{D}{D_{xx}}\right)^{2}   }\right]}{\frac{3\alpha}{4}\left[\frac{\hbar^{4}\tilde{Q}^{4}_{x}}{4m^{*2}}\left(\frac{D}{D_{xx}}\right)^{2}+\hbar^{2}v_{y}^{2}\tilde{Q}_{y}^{2}\left(\frac{D}{D_{yy}}\right)\right]^{-\frac{1}{4}}\times\hbar^{2}v^{2}_{y}\frac{D}{D_{yy}}\times 2\tilde{Q}_{y}
}.\nonumber\\
\end{eqnarray}
Let us ignore terms of order higher than \(\tilde{Q}_{x}^{2}\). Accordingly, Eq.~(\ref{eq:q-integral2}) reduces to
\begin{eqnarray}
\label{eq:q-integral3}
&&4\sum_{n^{(1)}_{min}}^{n^{(1)}_{max}}\int_{\ell^{-1}_{\phi}}^{\ell^{-1}_{e}}\frac{d\tilde{Q}_{x}}{(2\pi)^{2}D}\frac{4}{3\alpha}\frac{1}{2\hbar v
_{0}}\sqrt{\frac{D_{yy}}{D}}\times \left(\frac{1}{\tilde{Q}^{2}_{x}+\frac{D_{yy}}{\hbar^{2}v_{y}^{2}D}\left(n+\frac{1}{2}\right)^{\frac{4}{3}}\frac{1}{\alpha^{\frac{4}{3}}}}\right)\times \frac{1}{\left(n+\frac{1}{2}\right)^{\frac{1}{3}}\frac{1}{\alpha^{\frac{1}{3}}}} \nonumber\\
&&+4\sum_{n^{(2)}_{min}}^{n^{(2)}_{max}}\int_{0}^{\ell^{-1}_{\phi}}\frac{d\tilde{Q}_{x}}{(2\pi)^{2}D}\frac{4}{3\alpha}\frac{1}{2\hbar v
_{0}}\sqrt{\frac{D_{yy}}{D}}\times \left(\frac{1}{\tilde{Q}^{2}_{x}+\frac{D_{yy}}{\hbar^{2}v_{y}^{2}D}\left(n+\frac{1}{2}\right)^{\frac{4}{3}}\frac{1}{\alpha^{\frac{4}{3}}}}\right)\times \frac{1}{\left(n+\frac{1}{2}\right)^{\frac{1}{3}}\frac{1}{\alpha^{\frac{1}{3}}}}.
\end{eqnarray}
Note that 1) \(n^{(1)}_{\rm min}=0\), \(  n^{(1)}_{\rm max}=\alpha\left[\hbar^{2}v^{2}_{y}\left(\frac{D}{D_{yy}}\right)(\ell^{-2}_{e}-\tilde{Q}^{2}_{x})\right]^{\frac{3}{4}}-\frac{1}{2}\), and 2) \(n^{(2)}_{\rm min}=\alpha\left[\hbar^{2}v^{2}_{y}\left(\frac{D}{D_{yy}}\right)(\ell^{-2}_{\phi}-\tilde{Q}^{2}_{x})\right]^{\frac{3}{4}}-\frac{1}{2}\), \( n^{(2)}_{\rm max}=\alpha\left[\hbar^{2}v^{2}_{y}\left(\frac{D}{D_{yy}}\right)(\ell^{-2}_{e}-\tilde{Q}^{2}_{x})\right]^{\frac{3}{4}}-\frac{1}{2}\). 

\begin{figure}[htb]
\includegraphics[width=0.6\linewidth]{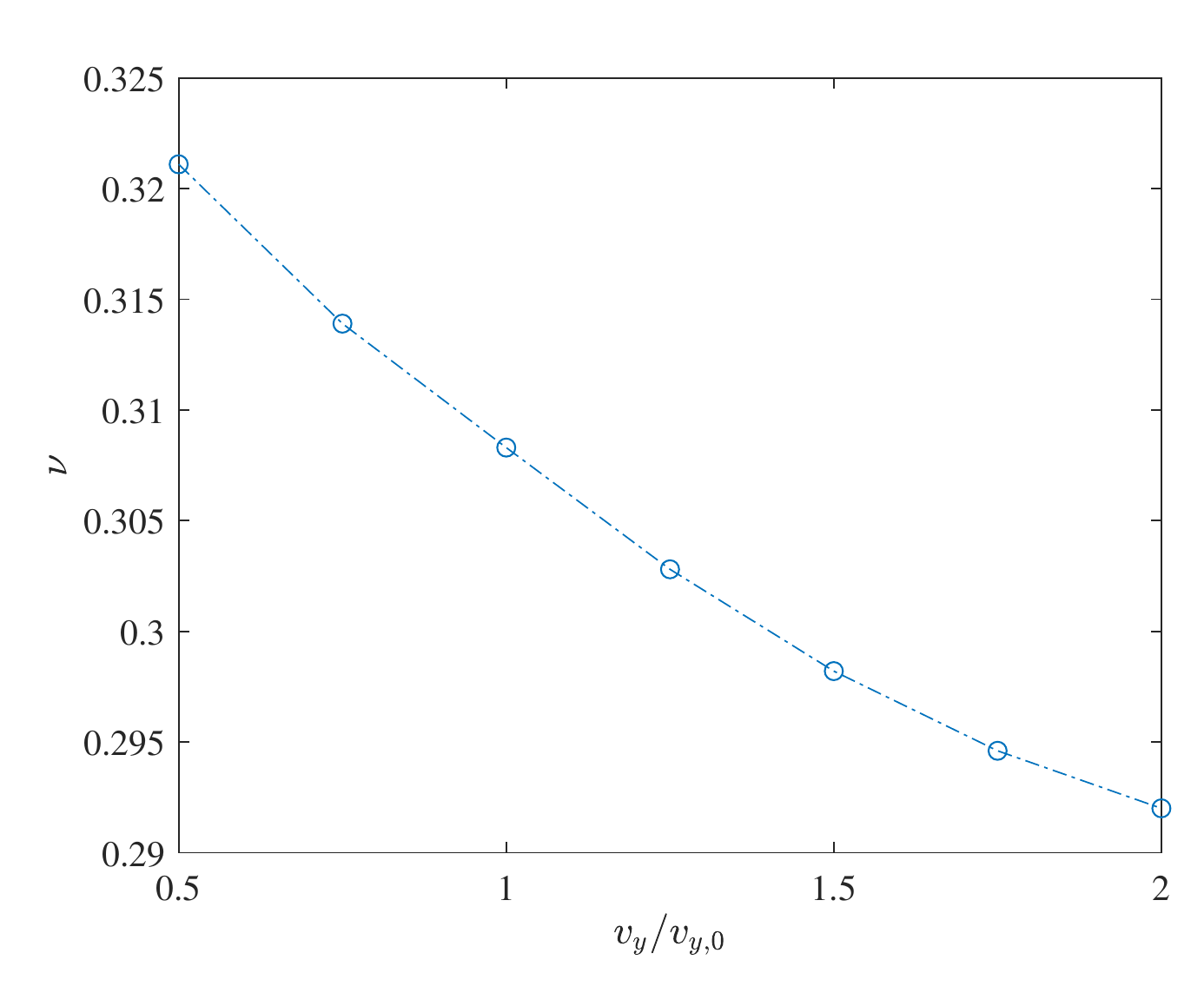}
\caption{The exponent \(\nu\) as a function of the system parameter \(v_{y}\). Here, referring to Jang {\em et al.} \cite{SM_Jang}, we set the unit velocity and effective mass at SDTP as \(v_{y,0}=1.2\times 10^{5}{\rm m}/{\rm s}\) and \(m^{*}=0.92m_{e}\), respectively. }
\label{fig:SM}
\end{figure}

We replace the \(n-\)sum by the \(n-\)integral as follows:
\begin{eqnarray}
\label{eq:q-integral4}
&&\frac{2\hbar v_{y}}{3\pi^{2}D}\sqrt{\frac{D}{D_{yy}}}\alpha^{\frac{2}{3}}\int_{\ell^{-1}_{\phi}}^{\ell^{-1}_{e}}d\tilde{Q}_{x}\int_{\frac{1}{2}}^{n^{(1)}_{max}+\frac{1}{2}}dn \frac{1}{n^{\frac{5}{3}}+\left(\alpha^{\frac{4}{3}}\tilde{Q}^{2}_{x}\frac{D}{D_{yy}}\hbar^{2}v^{2}_{y}\right)n^{\frac{1}{3}}}\nonumber\\
&&+\frac{2\hbar v_{y}}{3\pi^{2}D}\sqrt{\frac{D}{D_{yy}}}\alpha^{\frac{2}{3}}\int_{0}^{\ell^{-1}_{\phi}}d\tilde{Q}_{x}\int_{n^{(2)}_{min}+\frac{1}{2}}^{n^{(2)}_{max}+\frac{1}{2}}dn \frac{1}{n^{\frac{5}{3}}+\left(\alpha^{\frac{4}{3}}\tilde{Q}^{2}_{x}\frac{D}{D_{yy}}\hbar^{2}v^{2}_{y}\right)n^{\frac{1}{3}}}.
\end{eqnarray}
One can check the validity of this replacement by applying it to Eq.~(\ref{eq:quantization_dirac3}), giving the consistent result.
Now, let us introduce \(\gamma=\alpha^{\frac{4}{3}}\tilde{Q}^{2}_{x}\frac{D}{D_{yy}}\hbar^{2}v^{2}_{y}\). The \(n-\)integral can be computed by the following indefinite integral:
\begin{equation}
    \int dn \frac{1}{n^{\frac{5}{3}}+\gamma n^{\frac{1}{3}}}=\frac{3}{2\sqrt{\gamma}}\tan^{-1}\left({\frac{n^{\frac{2}{3}}}{\sqrt{\gamma}}}\right)+C.
\end{equation}
Accordingly, Eq.~(\ref{eq:q-integral4}) can be rewritten as
\begin{eqnarray}
\label{eq:51}
&&\frac{1}{\pi^{2}D}\left[\int_{0}^{\ell^{-1}_{e}}d\tilde{Q}_{x}\frac{1}{\tilde{Q}_{x}}\tan^{-1}\left({\frac{\sqrt{\ell^{-2}_{e}-\tilde{Q}^{2}_{x}}}{\tilde{Q}_{x}}}\right)-\int_{0}^{\ell^{-1}_{\phi}}d\tilde{Q}_{x}\frac{1}{\tilde{Q}_{x}}\tan^{-1}\left({\frac{\sqrt{\ell^{-2}_{\phi}-\tilde{Q}^{2}_{x}}}{\tilde{Q}_{x}}}\right)\right]\nonumber\\
&&-\frac{1}{\pi^{2}D}\int_{\ell^{-1}_{\phi}}^{\ell^{-1}_{e}}d\tilde{Q}_{x}\frac{1}{\tilde{Q}_{x}}\tan^{-1}\left({\frac{\alpha^{-\frac{2}{3}}}{2^{\frac{2}{3}}\tilde{Q}_{x}\hbar v_{y}}}\sqrt{\frac{D_{yy}}{D}}\right).
\end{eqnarray}
Since the first and second terms in Eq.~(\ref{eq:51}) are \(B-\)independent, the magnetoconductivity is only contributed by the third term. Finally, we obtain the ratio between the magnetoconductivity and the Boltzmann conductivity at the SDTP as
\begin{eqnarray}
\label{eq:magneto_semi-Dirac}
\frac{\Delta \sigma_{ii}(B)-\Delta \sigma_{ii}(0)}{\sigma^{\rm{B}}_{ii}}=\frac{1}{\pi^{3} D N_{0}\hbar}  \int_{\ell^{-1}_{\phi}}^{\ell^{-1}_{e}}d\tilde{Q}_{x}\frac{1}{\tilde{Q}_{x}}\tan^{-1}\left({\frac{\alpha^{-\frac{2}{3}}}{2^{\frac{2}{3}}\tilde{Q}_{x} \hbar v_{y}}}\sqrt{\frac{D_{yy}}{D}}\right).
\end{eqnarray}
Note that the ratio is also independent of the direction. Eq.~(\ref{eq:magneto_semi-Dirac}) indicates the \(B^{2/3}\) dependence of the magnetoconductivity in the weak field limit (\(\ell_{\phi}\ll \ell_{B}\)). As for the intermediate field regime (\(\ell_{e}\leq \ell_{B}\ll \ell_{\phi}\)), we predict that the magnetoconductivity will follow the power-law dependence on \(B\) with the exponent \(\nu\), which varies depending on the system parameter as shown in Fig.~\ref{fig:SM}. 

\section{Applicability of the WL theory to other anisotropic multiband and multivalley systems}

As for the multiband case, our results can be expanded to multiband systems straightforwardly by including the interference contribution from time-reversed paths in each band and associated Berry phase effect. As a concrete example, let us consider ABA stacked trilayer graphene in the presence of intravalley scattering. At a low Fermi energy where quantum corrections become important, this system has multiple dispersions near each valley: one linear dispersion with \(\pi\) Berry phase and one quadratic dispersion with \(2\pi\) Berry phase. Because an electron has its time-reversed path either within the same linear band or within the same quadratic band, one can compute the overall quantum corrections simply by including the interference contribution from the time-reversed paths in each band. According to our results, quantum corrections arising from linear (quadratic) dispersions with the Berry phase \(\pi\) (2\(\pi\)) would give WAL (WL) for intravalley scattering, following Eq.~(\ref{eq:ratio_dsm_intranode}) (Eq.~(\ref{eq:insulator_hikami_total2})). In this sense, our results could be expanded to compute the total quantum corrections in multiband systems.



\begin{thebibliography}{999}

\bibitem{Anderson1958}
P. W. Anderson, Absence of diffusion in certain random lattices, Phys. Rev. {\bf 109}, 1492 (1958). 

\bibitem{Lee1985}
P. A. Lee and T. V. Ramakrishnan, Disordered electronic systems, Rev. Mod. Phys. {\bf 57}, 287 (1985).

\bibitem{Datta1995}
S. Datta, \textit{Electronic Transport in Mesoscopic Systems} (Cambridge University Press, Cambridge, U.K., 1995).

\bibitem{Akkermans2007}
E. Akkermans and G. Montambaux, \textit{Mesoscopic Physics of Electrons and Photons} (Cambridge University Press, Cambridge, U.K., 2007).


\bibitem{Hikami1980}
S. Hikami, A. I. Larkin, and Y. Nagaoka, Spin-orbit interaction and magnetoresistance in the two dimensional random system, Prog. Theor. Phys. {\bf 63}, 707 (1980).

\bibitem{Altshuler1980}
B. L. Altshuler, D. Khmel'nitzkii, A. I. Larkin, and P. A. Lee, Magnetoresistance and Hall effect in a disordered two-dimensional electron gas, Phys. Rev. B {\bf 22}, 5142 (1984).


\bibitem{Wolfle1984}
P. W\"olfle and R. N. Bhatt, Electron localization in anisotropic systems, Phys. Rev. B {\bf 30}, 3542 (1984).

\bibitem{Beutler1988}
D. E. Beutler and N. Giordano, Localization and electron-electron interaction effects in thin Bi wires and films, Phys. Rev. B {\bf 38}, 8 (1988).

\bibitem{Bergmann1982}
G. Bergmann, Influence of spin-orbit coupling on weak localization, Phys. Rev. Lett. {\bf 48}, 1046 (1982).

\bibitem{Bergmann1984}
G. Bergmann, Weak localization in thin films: a time-of-flight experiment with conduction electrons, Phys. Rep.
{\bf107}, 1 (1984).




\bibitem{Suzuura2002}
H. Suzuura and T. Ando,
Crossover from Symplectic to Orthogonal Class in a Two-Dimensional Honeycomb Lattice,
Phys. Rev. Lett. {\bf 89}, 266603 (2002).

\bibitem{McCann2006}
E. McCann, K. Kechedzhi, V. I. Fal’ko, H. Suzuura, T. Ando, and B. L. Altshuler,
Weak-Localization Magnetoresistance and Valley Symmetry in Graphene,
Phys. Rev. Lett. {\bf 97}, 146805 (2006).

\bibitem{Gorbachev2007}
R. V. Gorbachev, F.V. Tikhonenko, A.S. Mayorov, D.W. Horsell, and A.K. Savchenko, Weak Localization in Bilayer Graphene, Phys. Rev. Lett. {\bf 98}, 176805 (2007) 

\bibitem{Kechedzhi2007}
K. Kechedzhi, V. I. Fal’ko, E. McCann, and B. L. Altshuler, 
Influence of Trigonal Warping on Interference Effects in Bilayer Graphene,
Phys. Rev. Lett. {\bf 98}, 176806 (2007).

\bibitem{Nomura2007}
K. Nomura, M. Koshino, and S. Ryu,
Topological Delocalization of Two-Dimensional Massless Dirac Fermions, Phys. Rev. Lett. {\bf 99}, 146806 (2007).


\bibitem{Imura2009}
K.-I. Imura, Y. Kuramoto, and K. Nomura, Weak localization properties of the doped 
\(Z_{2}\) topological insulator, Phys. Rev. B {\bf 80}, 085119 (2009).


\bibitem{Lu2015}
H.-Z. Lu and S.-Q. Shen, Weak antilocalization and localization in disordered and interacting Weyl semimetals, Phys. Rev. B {\bf 92}, 035203 (2015). 

\bibitem{Dai2016}
X. Dai, H.-Z. Lu, S.-Q. Shen, and H. Yao, Detecting monopole charge in Weyl semimetals via quantum interference transport, Phys. Rev. B {\bf 93}, 161110(R) (2016).


\bibitem{Lu2017}
H.-Z. Lu and S.-Q. Shen, Quantum transport in topological semimetals under magnetic fields, Front. Phys. {\bf 12}, 127201 (2017).


\bibitem{Chen2019}
W. Chen, H.-Z. Lu, and O. Zilberberg,
Weak Localization and Antilocalization in Nodal-Line Semimetals: Dimensionality and Topological Effects,
Phys. Rev. Lett. {\bf 122}, 196603 (2019).

\bibitem{Fu2019}
B. Fu, H.-W. Wang, and S.-Q. Shen, Quantum interference theory of magnetoresistance in Dirac materials, Phys. Rev. Lett. {\bf 122}, 246601 (2019).

\bibitem{He2011}
H.-T. He, G. Wang, T. Zhang, I.-K. Sou, G. K. L. Wong, and J.-N. Wang, Impurity effect on weak antilocalization in the topological insulator Bi\textsubscript{2}Te\textsubscript{3}, Phys. Rev. Lett. {\bf 106}, 166805 (2011).

\bibitem{Lu2011}
H.-Z. Lu, J. Shi, and S.-Q. Shen, Competition between weak localization and antilocalization in topological surface states, Phys. Rev. Lett. {\bf 107}, 076801 (2011).


\bibitem{Zhang2012}
H. B. Zhang, H. L. Yu, D. H. Bao, S. W. Li, C. X. Wang, and G. W. Yang, Weak localization bulk state in a topological insulator Bi\textsubscript{2}Te\textsubscript{3} film, Phys. Rev. B {\bf 86}, 075102 (2012).

\bibitem{Rodin2014}
A. S. Rodin, A. Carvalho, and A. H. Castro Neto, Strain-induced gap modification in black phosphorus, Phys. Rev. Lett. {\bf 112}, 176801 (2014).

\bibitem{Rudenko2014}
A. N. Rudenko and M. I. Katsnelson, Quasiparticle band structure and tight-binding model for single- and bilayer black phosphorus, Phys. Rev. B {\bf 89}, 201408 (2014).

\bibitem{Liu2015}
Q. Liu, X. Zhang, L. B. Abdalla, A. Fazzio, and A. Zunger, Switching a normal insulator into a topological insulator via electric field with application to phosphorene, Nano Lett. {\bf 15}, 1222 (2015).

\bibitem{Yuan2015}
S. Yuan, A. N. Rudenko, and M. I. Katsnelson, Transport and optical properties of single- and bilayer black phosphorus with defects, Phys. Rev. B {\bf 91}, 115436 (2015).

\bibitem{Kim2015}
J. Kim, S. S. Baik, S. H. Ryu, Y. Sohn, S. Park, B.-G. Park, J. Denlinger, Y. Yi, H. J. Choi, and K. S. Kim, Observation of tunable band gap and anisotropic Dirac semimetal state in black phosphorus, Science {\bf 349}, 723 (2015). 

\bibitem{Pereira2015}
J. M. Pereira, Jr. and M. I. Katsnelson, Landau levels of single-layer and bilayer phosphorene, Phys. Rev. B {\bf 92}, 075437 (2015).

\bibitem{Baik2015}
S. S Baik, K. S. Kim, Y. Yi, and H. J. Choi, Emergence of two-dimensional massless Dirac fermions, chiral pseudospins, and Berry’s phase in potassium doped few-layer black phosphorus, Nano Lett. {\bf 15} 7788 (2015).

\bibitem{Adroguer2016}
P. Adroguer, D. Carpentier, G. Montambaux, and E. Orignac, Diffusion of Dirac fermions across a topological merging transition in two dimensions, Phys. Rev. B {\bf 93}, 125113 (2016).

\bibitem{Yuan2016}
S. Yuan, E. van Veen, M. I. Katsnelson, and R. Rold\'an, Quantum Hall effect and semiconductor-to-semimetal transition in biased black phosphorus, Phys. Rev. B {\bf 93}, 245433 (2016).

\bibitem{Doh2017}
H. Doh and H. J. Choi, Dirac-semimetal phase diagram of two-dimensional black phosphorus, 2D Mater. {\bf 4}, 025071 (2017).

\bibitem{Kim2017}
J. Kim, S. S. Baik, S. W. Jung, Y. Sohn, S. H. Ryu, H. J. Choi, B.-J. Yang, and K. S. Kim, Two-dimensional Dirac fermions protected by space-time inversion symmetry in black phosphorus, Phys. Rev. Lett. {\bf 119}, 226801 (2017).

\bibitem{Park2019}
S. Park, S. Woo, and H. Min,
Semiclassical Boltzmann transport theory of few-layer black phosphorous in various phases,
2D Mater. {\bf 6}, 025016 (2019).

\bibitem{Jang2019}
J. Jang, S. Ahn, and  H. Min, Optical conductivity of black phosphorus with a tunable  electronic structure, 2D Mater. {\bf 6} ,025029 (2019).

\bibitem{Li2014}
L. Li, Y. Yu, G. J. Ye, Q. Ge, X. Ou, H. Wu, D. Feng, X. H. Chen, and Y. Zhang, Black phosphorus field-effect transistors, Nat. Nanotechnol. {\bf 9}, 372 (2014).


\bibitem{Xia2014}
F. Xia, H. Wang, and Y. Jia, Rediscovering black phosphorus as an anisotropic layered material for optoelectronics and electronics, Nat. Commun. {\bf 5} 4458 (2014).

\bibitem{Tran2014}
V. Tran, R. Soklaski, Y. Liang, and L. Yang, Layer-controlled band gap and anisotropic excitons in few-layer black phosphorus, Phys. Rev. B {\bf 89}, 235319 (2014).

\bibitem{Qiao2014}
J. Qiao, X. Kong, Z.-X. Hu, F. Yang, and W. Ji, High-mobility transport anisotropy and linear dichroism in few-layer black phosphorus, Nat. Commun. {\bf 5}, 4475 (2014).

\bibitem{Xiang2015}
Z. J. Xiang, G. J. Ye, C. Shang, B. Lei, N. Z. Wang, K. S. Yang, D. Y. Liu, F. B. Meng, X. G. Luo, L. J. Zou, Z. Sun, Y. Zhang, and X. H. Chen, Pressure-induced electronic transition in black phosphorus, Phys. Rev. Lett. {\bf 115}, 186403 (2015).

\bibitem{Du2016}
Y. Du, A. T. Neal, H. Zhou, and P. D. Ye, Weak localization in few-layer black phosphorus, 2D Mater. {\bf 3}, 024003 (2016).

\bibitem{Shi2016}
Y. Shi, N. Gillgren, T. Espiritu, S. Tran, J. Yang, K. Watanabe, T. Taniguchi, and C. N. Lau, Weak localization and electron–electron interactions in few layer black phosphorus devices, 2D Mater. {\bf 3}, 034003 (2016).

\bibitem{Hemsworth2016}
N. Hemsworth, V. Tayari, F. Telesio, S. Xiang, S. Roddaro, M. Caporali, A. Ienco, M. Serrano-Ruiz, M. Peruzzini, G. Gervais, T. Szkopek, and S. Heun, Dephasing in strongly anisotropic black phosphorus, Phys. Rev. B {\bf 94}, 245404 (2016).

\bibitem{Li2017}
C.-H. Li, Y.-J. Long, L.-X. Zhao, L. Shan, Z.-A. Ren, J.-Z. Zhao, H.-M. Weng, X. Dai, Z. Fang, C. Ren, and G.-F. Chen, Pressure-induced topological phase transitions and strongly anisotropic magnetoresistance in bulk black phosphorus, Phys. Rev. B {\bf 95}, 125417 (2017).


\bibitem{Mahan2000}
G. D. Mahan, \textit{Many-particle physics}, 3rd ed. (Springer, Berlin, 2000).


\bibitem{Coleman2016} 
P. Coleman, \textit{Introduction to Many-Body Physics} (Cambridge University Press, Cambridge, U.K., 2016).

\bibitem{Ashcroft1976}
N. W. Ashcroft and N. D. Mermin, \textit{Solid State Physics} (Brooks Cole, Pacific Grove, CA, 1976).

\bibitem{Park2017}
S. Park, S. Woo, E. J. Mele, and H. Min,
Semiclassical Boltzmann transport theory for multi-Weyl semimetals,
Phys. Rev. B {\bf 95}, 161113(R) (2017).

\bibitem{Kim2019}
S. Kim, S. Woo, and H. Min,
Vertex corrections to the dc conductivity in anisotropic multiband systems,
Phys. Rev. B {\bf 99}, 165107 (2019).

\bibitem{see SM}
See Supplemental Material for the detailed derivations of the Cooperon ansatz, WL and WAL corrections, magnetoconductivity, and the discussion on the applicability of the generalized WL theory to other anisotropic multiband and multivalley systems. 

\bibitem{Dietl2008}
P. Dietl, F. Pi\'echon, and G. Montambaux, New magnetic field dependence of Landau levels in a graphenelike structure, Phys. Rev. Lett. {\bf 100}, 236405 (2008). 

\bibitem{Dyson1962}
F. J. Dyson, Statistical theory of the energy levels of complex systems. \RNum{1}, J. Math. Phys. (N.Y.) {\bf 3}, 140 (1962).


\end{thebibliography}

\begin{thebibliography}{999}

\bibitem{SM_Park2019}
Sanghyun Park, Seungchan Woo, and Hongki Min,
Semiclassical Boltzmann transport theory of few-layer black phosphorous in various phases,
2D Mater. {\bf 6}, 025016 (2019).


\bibitem{SM_Mahan}
Gerald D. Mahan, \textit{Many-particle physics}, 3rd ed. (Springer, Berlin, 2000).


\bibitem{SM_Coleman} 
P. Coleman, \textit{Introduction to Many-Body Physics} (Cambridge University Press, Cambridge, 2016).

\bibitem{SM_Ashcroft}
N. W. Ashcroft and N. D. Mermin, \textit{Solid State Physics} (Brooks Cole, Pacific Grove, CA, 1976).

\bibitem{SM_Kim}
Sunghoon Kim, Seungchan Woo, and Hongki Min,
Vertex corrections to the dc conductivity in anisotropic multiband systems,
Phys. Rev. B {\bf 99}, 165107 (2019).

\bibitem{SM_Suzuura}
Hidekatsu Suzuura and Tsuneya Ando,
Crossover from Symplectic to Orthogonal Class in a Two-Dimensional Honeycomb Lattice,
Phys. Rev. Lett. {\bf 89}, 266603 (2002).

\bibitem{SM_Lu}
Hai-Zhou Lu and Shun-Qing Shen, Weak antilocalization and localization in disordered and interacting Weyl semimetals, Phys. Rev. B {\bf 92}, 035203 (2015). 

\bibitem{SM_Chen}
Wei Chen, Hai-Zhou Lu, and Oded Zilberberg,
Weak Localization and Antilocalization in Nodal-Line Semimetals: Dimensionality and Topological Effects,
Phys. Rev. Lett. {\bf 122}, 196603 (2019).

\bibitem{SM_Pereira}
J. M. Pereira, Jr. and M. I. Katsnelson, Landau levels of single-layer and bilayer phosphorene, Phys. Rev. B {\bf 92}, 075437 (2015).

\bibitem{SM_Dietl}
Petra Dietl, Fr\'ed\'eric Pi\'echon, and Gilles Montambaux, New magnetic field dependence of Landau levels in a graphenelike structure, Phys. Rev. Lett. {\bf 100}, 236405 (2008). 

\bibitem{SM_Jang}
Jiho Jang, Seongjin Ahn, and  Hongki Min, Optical conductivity of black phosphorus with a tunable  electronic structure, 2D Mater. {\bf 6} ,025029 (2019).

\end{thebibliography}
\end{document}